\documentclass[]{aa_nolob}

\usepackage{txfonts}
\usepackage{natbib}
\usepackage{graphicx}
\usepackage[figuresright]{rotating}

\bibpunct{(}{)}{;}{a}{}{,}     
\newcommand{\pdx}[2]{#1\times10^{#2}}

\begin{document}

\title{C2D Spitzer-IRS spectra of disks around T Tauri stars}
\subtitle{II. PAH emission features}
\titlerunning{PAH emission features from disks}

\author{
V.C. Geers \inst{1}
 \and J.-C. Augereau \inst{1,2}
 \and K.M. Pontoppidan \inst{1,3 }
 \and C.P. Dullemond \inst{4}
 \and R. Visser \inst{1}
 \and J.E. Kessler-Silacci \inst{5}
 \and N.J. Evans, II\inst{5}
 \and E.F. van Dishoeck \inst{1}
 \and G.A. Blake \inst{3}
 \and A.C.A. Boogert \inst{6}
 \and J.M. Brown \inst{3}
 \and F. Lahuis \inst{1,7}
 \and B. Mer\'{\i}n \inst{1}
}

\offprints{V.C. Geers; \email{vcgeers@strw.leidenuniv.nl}}

\institute{Leiden Observatory, P.O. Box 9513, 2300 RA Leiden, The Netherlands
\and Laboratoire d'Astrophysique de l'Observatoire de Grenoble, B.P. 53, 38041 Grenoble Cedex 9, France
\and Division of GPS, Mail Code 150-21, California Institute of Technology, Pasadena, CA 91125, USA
\and Max-Plank-Institut f\"{u}r Astronomie, Koenigstuhl 17, 69117 Heidelberg, Germany
\and Department of Astronomy, University of Texas, 1 University Station C1400, Austin, TX 78712-0259, USA
\and Division of PMA, Mail Code 105-24, California Institute of Technology, Pasadena, CA 91125, USA
\and SRON, P.O. Box 800, 9700 AV Groningen, The Netherlands
}
\authorrunning{Geers et al.}

\date{Received $<$date$>$; Accepted $<$date$>$}

\abstract
{}
{We search for Polycyclic Aromatic Hydrocarbon (PAH) features towards
young low-mass (T Tauri) stars and compare them
with surveys of intermediate mass (Herbig Ae/Be)
stars. The presence and strength of the PAH features are interpreted with
disk radiative transfer models exploring the PAH feature dependence on
the incident UV radiation, PAH abundance and disk parameters.}
{Spitzer Space Telescope 5--35 $\mu$m spectra of 54 pre-main
sequence stars with disks were obtained, consisting of
38 T Tauri, 7 Herbig Ae/Be and 9 stars with unknown spectral type.}
{Compact PAH emission is detected towards at least 8 sources of which
5 are Herbig Ae/Be stars.  The 11.2 $\mu$m PAH feature is detected
in all of these sources, as is the 6.2 $\mu$m PAH feature for the 4
sources for which short wavelength data are available. However, the
7.7 and 8.6 $\mu$m features appear strongly in only 1 of these 4
sources. Based on the 11.2 $\mu$m feature, PAH emission is observed
towards at least 3 T Tauri stars, with 14 tentative detections,
resulting in a lower limit to the PAH detection rate of 8\,\%.  The
lowest mass source with PAH emission in our sample is \object{T Cha}
with a spectral type G8. 
All 4 sources in our sample with evidence for dust holes 
in their inner disk show PAH emission, increasing the feature/continuum ratio.
Typical 11.2 $\mu$m line intensities are
an order of magnitude lower than those observed for the more massive
Herbig Ae/Be stars. Measured line fluxes indicate PAH abundances that
are factors of 10--100 lower than standard interstellar values.
Conversely, PAH features from disks exposed to stars with $T_{\rm
eff}\leq 4200$~K without enhanced UV are predicted to be
below the current detection limit, even for high PAH abundances.  Disk
modeling shows that the 6.2 and 11.2 $\mu$m features are the best
PAH tracers for T Tauri stars, whereas the 7.7 and 8.6 $\mu$m bands
have low feature over continuum ratios due to the strongly rising
silicate emission.}
{}


\maketitle

\section{Introduction}
Polycyclic Aromatic Hydrocarbons (PAHs) have been observed in a wide
variety of sources in our own and external galaxies. Within the Milky Way PAHs are observed in the 
diffuse medium, dense molecular clouds, circumstellar envelopes, 
and (proto-)planetary nebulae (see \citealt{pee04} for a summary).
A common characteristic of all of these sources is that they are exposed 
to copious ultraviolet (UV) photons.
The UV radiation drives the molecules into excited electronic states,
which subsequently decay to lower electronic states through a non-radiative
process called internal conversion, followed by vibrational emission
in the available C-H and C-C stretching and bending vibrational modes
at $3.3$, $6.2$, $7.7$, $8.6$, $11.2$, $12.8$ and 16.4 $\mu$m.  Thus, PAH
molecules form an important diagnostic of UV radiation.

In recent years, PAH emission has also been detected from disks around
young stars in ground-based and {\it Infrared Space Observatory} (ISO)
spectra \citep{ker00,hon01,pee02,boe03,prz03,ack04}.  
The PAH emission is thought to originate from
the surface layer of a (flaring) disk exposed to radiation from the central 
star \citep[cf. models by][ hereafter
H04]{man99,hab04}. Indeed, ground-based spatially resolved
observations show that the features come from regions with sizes
typical of that of a circumstellar disk (radius $<$12 AU at 3.3
$\mu$m, $<$100 AU at 11.2 $\mu$m) \citep{gee04,boe03,hab05}.  Searches for
PAHs in disks are important because in addition to being a tracer of
the strength of the stellar radiation field and disk geometry, the PAHs also affect
the disk structure and chemistry. For example, the high opacity of PAHs at FUV
wavelengths \citep{mat05a} could significantly reduce the stellar UV radiation in the
inner disk while photoionization of PAHs produces energetic
electrons which are a major heating mechanism for the gas in the upper
layers of the disk where the gas and dust temperatures are not well
coupled \citep{jon04,kam04b}.

Detections of PAH features also provide diagnostics of the presence of
small grains in the surface layers of disks and the dust evolution
through grain growth and dust settling. Evidence for grain growth has
been found from modeling of the silicate features from Herbig Ae/Be (hereafter HAeBe) 
\citep{boe03} and T Tauri disks (\citealp{prz03}; \citealp{kes05}, 2006) 
and more indirectly from the modeling of the observed H$_2$
emission features from T Tauri disks \citep{ber04}.  
PAHs are considered to be on the small end of the size distribution of grains.
An important question is whether PAHs have a different timescale for settling 
and/or growth compared to that of larger silicate/carbon dust grains. 

So far, most data obtained on PAHs refer to 
relatively bright features in the spectra of intermediate mass HAeBe stars.
A recent ISO spectroscopic survey has detected PAH features, in
particular the most frequently observed 6.2\,$\mu$m feature, towards
57\,\% of a sample of 46 HAeBe stars \citep{ack04}. \citet{mee01}
classified the ISO observed spectral energy distributions (SEDs) of
these intermediate mass young stars into two groups (group I and II). \citet{dom03}
interpreted these SEDs in the context of a passive disk model with a
puffed-up inner rim \citep{dul01} and proposed that group I sources
with larger mid-infrared excess have flaring disks while group II
sources are consistent with small and/or self-shadowed disks.
\citet{ack04} showed that the group I sources with strong mid-infrared
relative to near-infrared excess display significantly more PAH
emission than the group II sources with weaker mid-infrared excesses,
consistent with the idea that the PAH emission originates mostly from
the disk surface.

The arrival of the Spitzer Space Telescope \citep{wer04} with the
InfraRed Spectrograph (IRS) \citep{hou04} provides the opportunity to
extend these studies to disks around fainter low mass T Tauri
stars. For sources of spectral type G and later, the stellar UV field
is orders of magnitude weaker than for HAeBe stars, which will
directly affect the PAH excitation and emission. On the other hand,
enhanced UV radiation has been detected for some T Tauri stars (e.g.,
Costa et al.\ 2000, Bergin et al.\ 2003).
Such an enhanced UV field should be directly reflected
in the intensity of the PAH features if these molecules are present in
normal abundances. Furthermore, ionized PAHs can be excited by less energetic, optical photons from these cooler stars \citep{mat05a}.

In this paper we present detections of PAH features toward T Tauri
stars from our initial set of Spitzer IRS 5-38 $\mu$m spectroscopic
observations that were taken as part of the Spitzer Legacy program
``From Molecular Cores to Planet-Forming Disks'' \citep{eva03}
(hereafter c2d). The c2d targets consist of a large number of sources
with infrared excess in five of the nearest large star-forming
regions: Chamaeleon, Lupus, Ophiuchus, Perseus and Serpens. In Paper
I, the silicate 10 and 20 $\mu$m features from pre-main sequence stars
with disks are presented and analyzed (Kessler-Silacci et al.\
2006). The data are used here to search for PAH features and address a
number of outstanding questions. For how many low mass stars can PAH
features be found and how does this compare to HAeBe stars? Can limits
be put on the abundance of PAHs and thus indirectly on that of the
smallest grains?  Which factors influence the appearance of PAH
features in disks?  Can we quantify any additional UV or optical radiation from
the strength of the PAH features?

Section \ref{sec:obs} describes the sample selection, observations and
reduction method. The results for the observed PAH features are
presented in Sect.\,\ref{sec:results} and discussed in Sects.\,\ref{sec:pahfeatures} and \ref{sec:pahindisks}.  Conclusions are
presented in Sect.\,\ref{sec:conclusions}.

\section{Observations and data reduction}
\label{sec:obs}
Mid-infrared spectra were obtained for intermediate and low mass stars
with circumstellar disks with the IRS aboard Spitzer as part of the
c2d Legacy program. All targets were observed with the Short-High (SH)
module (spectral resolving power of $R\sim600$) covering the
wavelength range 10--20 $\mu$m and thereby potential PAH features at
11.2 and 12.8\,$\mu$m. For a fraction of our sources, Short-Low (SL)
observations ($R\sim100$, 5--14.5\,$\mu$m) were obtained as well,
which cover the 6.2, 7.7, 8.6, 11.2 and 12.8\,$\mu$m PAH
features. For the remainder of our sources, the SL observations are in
the Guaranteed Time Observation (GTO) IRS program (J.R.\ Houck) and 
were at the time of paper submission not yet available.

The original sample selection of the c2d program is described in \citet{eva03}. 
In our sample of sources observed to date there are 7 HAeBe stars, 38
T Tauri stars and 9 sources without known spectral types. These include 
the 40 T Tauri stars and 7 HAeBe stars in Kessler-Silacci et al.\ (2006), 
although we label 4 of their T Tauri sources here as ``unclassified'' for lack of 
spectral type (\object{RNO 15}, \object{IRAS 03446+3254}, \object{VSSG1}, 
\object{CK 4}) and we label 1 of their HAeBe stars here as a T Tauri star 
since it has spectral type M6 (\object{DL Cha}). In addition, we have 
added the c2d IRS spectra taken for 1 HAeBe star (\object{HD101412}), 
1 T Tauri star, \object{Sz 84}, and 5 sources without known spectral 
types (\object{EC 69},  \object{EC 88}, \object{EC 90}, \object{EC 92}, 
\object{RNO 91}). Our current sample of 54 sources is therefore comparable 
in size to that of the HAeBe
stars observed with ISO. We grouped the sample sources in intermediate
and low mass young stars following \citet{the94}, as HAeBe
stars with spectral types of F7 and earlier, and T\,Tauri stars with
spectral type F8 and later.  A summary of the properties of all disk
sources can be found in \citet[ Paper I]{kes06}
and Mer\'{\i}n et al.\ (in prep.). The sample includes the spectrum of an off-position, taken toward the Ophiuchus cloud at a position devoid of infrared sources, to compare with a typical interstellar 
medium PAH spectrum.

All spectra were extracted from the SSC pipeline version S12.0.2
BCD images, using the c2d reduction pipeline (Kessler-Silacci et al. 2006 and
Lahuis, in preparation).
The processing includes bad-pixel correction through interpolation
using a source profile fit in the cross-dispersion direction.
The source profile fitting also gives an estimate of the local sky
contribution in the high-resolution spectra.
The extracted spectra are defringed using the IRSFRINGE package
developed by the c2d team and individual orders are in some cases
corrected by small scaling corrections ($\lesssim$\,5\,\%) to match the order with the shortest wavelength. Sky subtraction is applied and in a few cases
the SH module is scaled down ($\lesssim$\,5\,\%) in flux to match the SL spectrum.

\section{T\,Tauri stars with PAH features}
\label{sec:results}
%
\begin{table*}
\begin{center}
\caption{Sources with potential PAH emission and their characteristics}
\label{tbl:obssum}
\begin{tabular}{lllllllll}
\hline\hline
\noalign{\smallskip}
Name                 & RA (J2000) & Dec (J2000) & AOR key & Modules$^{\mathrm{a}}$ &  Dist. [pc] & Sp. Type & H$\alpha$ ($\AA$) & Ref \\
\hline
\noalign{\smallskip}
\object{LkH$\alpha$ 330} & 03 45 48.3 & $+$32 24 12 & 0005634816 & SL LL1 SH LH &  250 & G3e $^\mathrm{e}$& 11.4$^{\mathrm{b}}$--20.3 & 
3; 
4,5; 
4,5\\ 
%
\object{RR Tau}          & 05 39 30.5 & $+$26 22 27 & 0005638400 & SL LL1 SH LH &  160 & A0e--A0IVe $^\mathrm{e}$ & 21.2 & 
11; 
12,13; 
5\\ 
\object{HD 98922}        & 11 22 31.7 & $-$53 22 12 & 0005640704 & SH LH        &  $>$540$^\mathrm{f}$ & B9Ve $^\mathrm{e}$ & 27.9 & 
1; 
17; 
10\\ 
%
\object{HD 101412}       & 11 39 44.5 & $-$60 10 28 & 0005640960 & SL SH LH     &  160 & B9.5Ve $^\mathrm{e}$ & 20.4 & 
10; 
16; 
10\\ 
%
\object{T Cha}           & 11 57 13.5 & $-$79 21 32 & 0005641216 & SH LH        &  66$^{+19}_{12}$ & G8e $^\mathrm{e}$ & 2--10 & 
1; 
2; 
2,18\\ 
%
\object{HD 135344}       & 15 15 48.4 & $-$37 09 16 & 0005657088 & SH LH        &  140& F4Ve & 17.4  & 
8; 
9; 
10\\ 
%
\object{EM* SR 21 N}       & 16 27 10.3 & $-$24 19 13 & 0005647616 & SH LH        &  125 & G2.5 & 0.54 & 
6; 
7; 
19 \\ 
%
\object{VV Ser}          & 18 28 47.9 & $+$00 08 40 & 0005651200 & SL SH LH     &  259 & A0Ve & 22--51$^{\mathrm{c}}$--81.3 & 
14; 
13; 
15,4,10\\ 
%
\hline
\noalign{\smallskip}
Off-position  & 16 24 00       & $-$24 00 00 & 0005654272 & SH LH & -- & -- &  -- & -- \\
\object{VSSG 1} $^{\mathrm{d}}$         & 16 26 18.9 & $-$24 28 20 & 0005647616 & SH LH        &  125 & \ldots & \ldots & 
6; 
 -;
 - \\
\object{Haro 1--17} $^{\mathrm{d}}$     & 16 32 21.9 & $-$24 42 15 & 0011827712 & SL LL1 SH LH &  125 & M2.5e $^\mathrm{e}$& 15 & 
6; 
2; 
2\\ 
%
\hline
\end{tabular}
\begin{list}{}{}
\item $^{\mathrm{a}}$  SL = SL1 + SL2
\item $^{\mathrm{b}}$ average from 3 measurements
\item $^{\mathrm{c}}$ average from 19 measurements
\item $^{\mathrm{d}}$ PAH feature detected in background spectrum, not associated with the source
\item $^{\mathrm{e}}$ label `e' added to spectral type here, based on H$\alpha > 0$, not taken from reference.
\item $^{\mathrm{f}}$ distance uncertain
\item References for distance; spectral type; H$\alpha$: 
1: \citet{anc98}, 
2: \citet{alc93}, 
3: Enoch et al. 2005, 
4: \citet{fer95},
5: \citet{coh79}, 
6: Assumed distance to Oph cloud \citep{geu89}, 
7: \citet{pra03}, 
8: \citet{ack04}, 
9: \citet{dun97}, 
10: \citet{ack05}, 
11: Assumed distance to Taurus-Auriga cloud \citet{ken94}, 
12: \citet{her04}, 
13: \citet{mor01}, 
14: \citet{str96}, 
15: \citet{fin84}, 
16: \citet{the94}, 
17: \citet{hou78},
18: \citet{alc95},
19: \citet{mar98}.
\item Note: sources in the bottom portion of the Table 
have spectra with PAH features that are fully attributed to background emission.
\end{list}
\end{center}
\end{table*}
\begin{figure}
  \centering
  \includegraphics[width=\columnwidth]{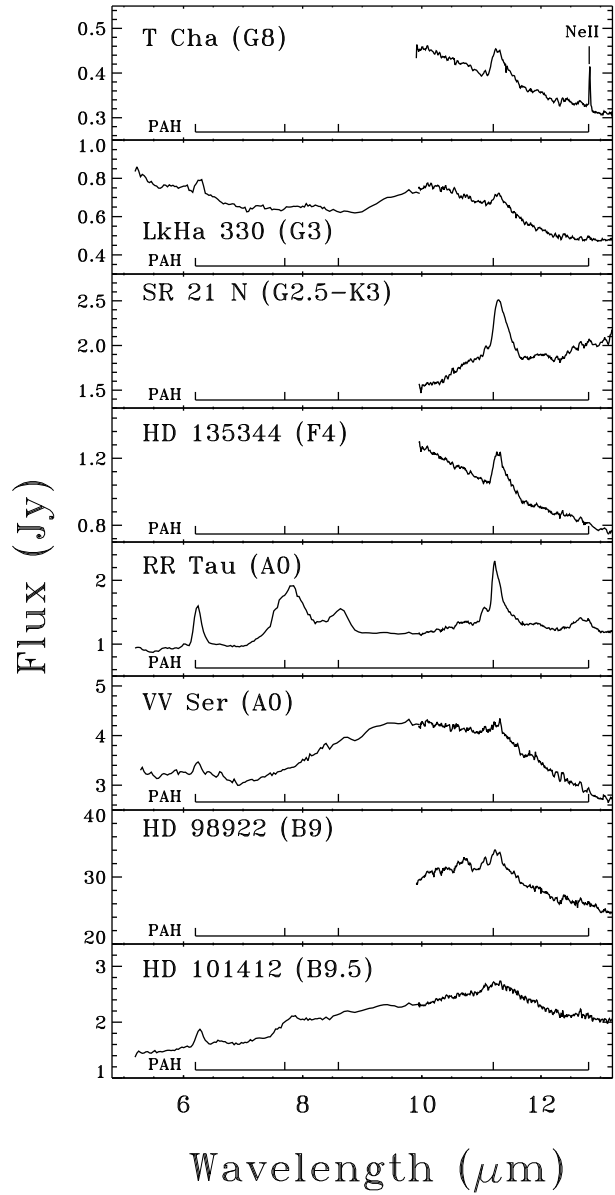}
  \caption{Spitzer IRS spectra of sources with PAH features, comprised of the SL (5--10 $\mu$m) and SH (10--20 $\mu$m) modules.
  The location of PAH features is indicated with markers at $6.2$, $7.7$, $8.6$, $11.2$ and $12.8$\,$\mu$m.}
  \label{fig:allspecs}
\end{figure}
\begin{figure}
  \centering
  \includegraphics[width=\columnwidth]{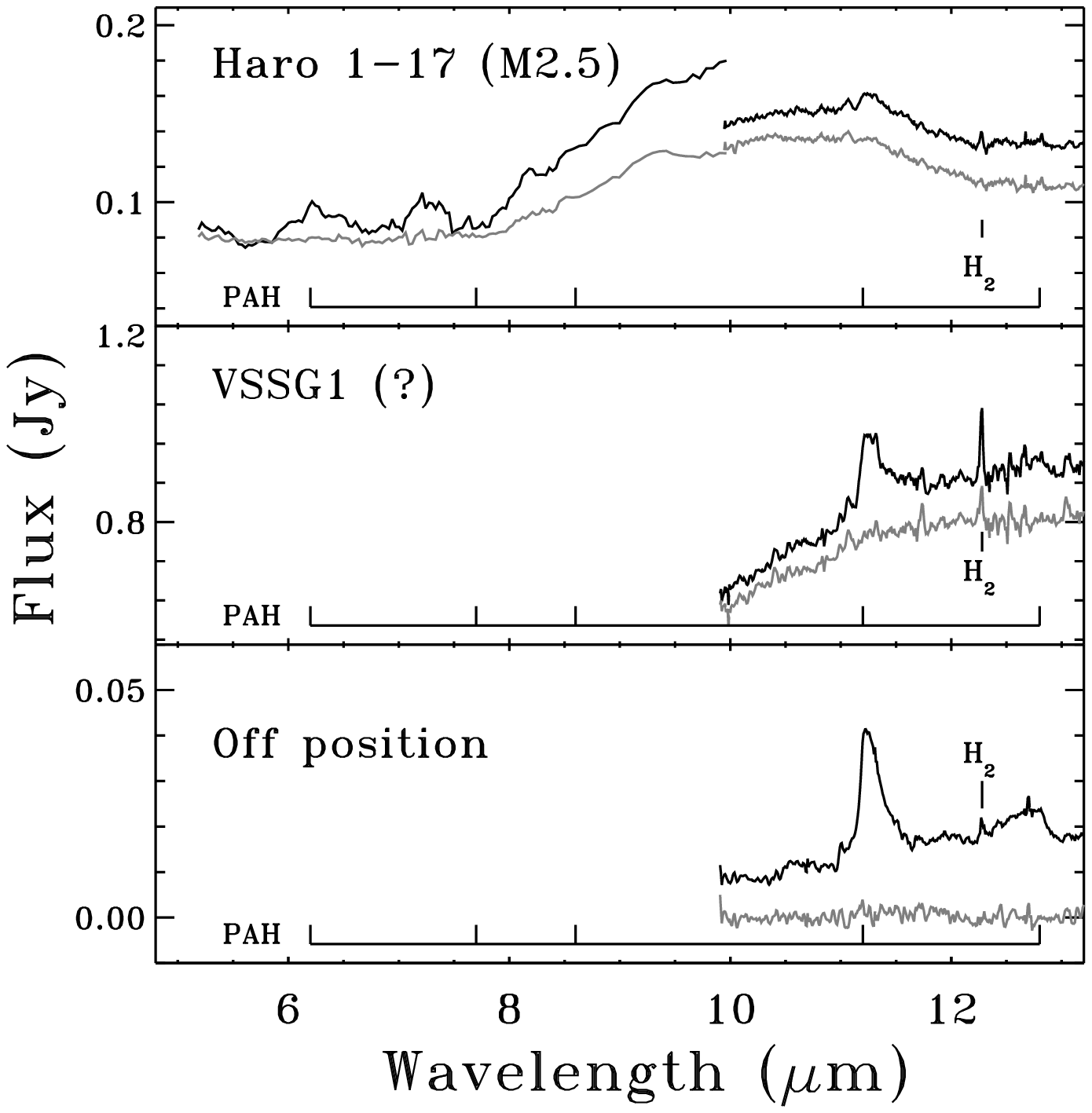}
  \caption{Spitzer IRS spectra (black line) of sources with PAH features that are associated with extented cloud emission, comprised of the SL (5--10 $\mu$m) and SH (10--20 $\mu$m) modules. The extended emission corrected spectrum is shown in grey.
  The location of PAH features is indicated with markers at $6.2$, $7.7$, $8.6$, $11.2$ and $12.8$\,$\mu$m.}
  \label{fig:allspecs_bgpahs}
\end{figure}

\subsection{Identification of PAH features}
Among the $54$ disk sources observed to date, a relatively small sample of $8$ sources shows one or more clear emission feature that we attribute to PAHs
(Fig.~\ref{fig:allspecs}). 
Table~\ref{tbl:obssum} summarizes the characteristics of these sources. 
They consist of $3$ T\,Tauri stars and $5$ HAeBe stars. Besides the PAH features, some sources show the H$_{2}$ S(2) line, whereas the [N$_{\mathrm{e}}$II] 12.8 $\mu$m line is detected toward T Cha. Unlabelled narrow features are likely spurious. 

Disk sources with PAH emission features were identified as follows.
When both SL and SH spectra are available, we required detection of
both a clear $11.2$\,$\mu$m feature and a clear $6.2$, $7.7$ and/or
$8.6$\,$\mu$m feature.  For sources where only a SH spectrum
is available ($\lambda > 10\,$$\mu$m), the identification is more
critical and largely relies on both the presence of a clear feature at
$11.2$\,$\mu$m and its shape. We compared potential $11.2$\,$\mu$m features
with that observed in the off-position spectrum and selected the sources with the lowest residuals, see Sect.\,\ref{sec:featcomp}. 

This method can introduce a bias against sources with mixed crystalline silicate and PAH features, because crystalline forsterite has a characteristic feature at almost the same
central wavelength as the $11.2$\,$\mu$m PAH feature. To distinguish
between these two possible assignments, we searched for other
crystalline silicate features at wavelengths longer than
$11.2$\,$\mu$m (e.g. $16.2$, $18.9$, $23.7$ and
$33.6$\,$\mu$m). The large wavelength coverage out to 35\,$\mu$m is
a significant advantage compared with ground-based data. 
Three of the five HAeBe and 2 of the 3 T Tauri sources show (tentative) evidence 
for crystalline silicates at longer wavelengths, indicated in Table~\ref{tbl:lineflux} 
(Kessler-Silacci et al.\ 2006). 
In these sources, a contribution from crystalline silicates to the observed $11.2$--$11.3$ 
$\mu$m feature cannot be excluded. The sources illustrate well the difficulties inherent in
the identification of PAH emission features at $11.2\,$$\mu$m when
lacking any information on the presence/absence of other
characteristic PAH features at shorter wavelengths. We note also that the $11.2$\,$\mu$m PAH feature can be blended with the broad amorphous silicate feature whose strength
and spectral width vary with grain size. This may have led us to miss
sources with weak 11.2\,$\mu$m emission from PAHs (see discussion in Sect.\,\ref{sec:stat}), although the characteristic shape of the 11.2\,$\mu$m PAH feature can 
help in distinguishing between PAHs and silicates, as discussed in Sect.\,\ref{sec:featcomp}.

Finally, sources with PAH features but showing silicate and/or ice
features in absorption have been excluded from the sample presented
here.  For comparison, Fig.~\ref{fig:allspecs_bgpahs} includes
observations of the off-source spectrum and of two more late type
sources, \object{Haro 1-17} and \object{VSSG1} which were selected on
the above mentioned criteria but for which background extraction shows that the 
PAH emission features are fully associated with the extended cloud 
emission (see below).

\subsection{Spatial extent of PAH emission}
\label{ssec:spatextent}
An important question is whether the PAH emission originates primarily
from the star+disk system or from an extended nebulosity around the
star. To properly answer this question, either spatially resolved
spectra or images in specific PAH band filters (both at feature
wavelength and slightly off-peak for determination of strength and
extent of continuum emission) with sufficient (subarcsec) spatial
resolution are required.  Since such data are lacking for our sample
we use the extracted background and long-slit spectra to provide
constraints on the extent of the emission.

\begin{figure}
  \centering
  \includegraphics[width=\columnwidth]{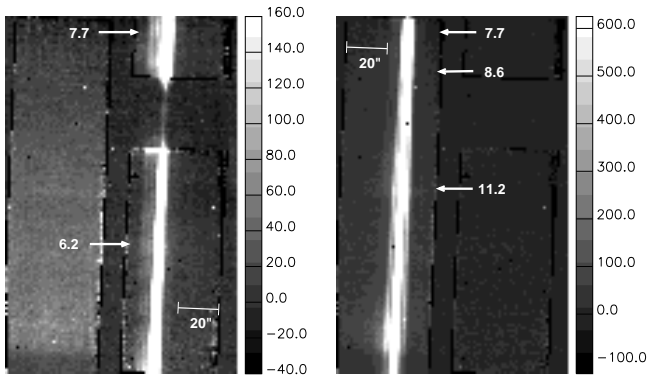}
  \caption{2D spectral image of modules SL2 (left panel) and SL1 (right panel) for \object{RR Tau}. The quantity shown is the ADU in units of $\mathrm{e^{-} s^{-1}}$. The positions of the 6.2, 7.7, 8.6 and 11.2 $\mu$m PAH features are indicated with labels. The horizontal bar at bottom right of the left panel and top left of the right panel indicates a spatial extent of 20$''$.}
  \label{fig:rrtau_sl_2dspec}
\end{figure}

The SL spectra are taken using long-slit spectroscopy and the 2D
spectral images can be used to determine if the features are seen 
extended along the entire width of the slit. An example 2D spectral image 
of module SL1 is shown for \object{RR Tau} in Fig.~\ref{fig:rrtau_sl_2dspec}, 
where no extended emission is seen along the slit. The pixel size of the 
SL module is $1.8''$, which means that any feature originating from a 
region smaller than 2 pixels (e.g. $540\,$ AU at 150\,pc) will be spatially
unresolved. A similar lack of extended emission in SL1 is seen for all of 
our PAH sources with the exception of \object{Haro 1-17}. Thus, the PAH 
emission for these sources is constrained to originate from a 
region of at most 3.6$''$. 

The SL spectra presented have been corrected for the estimated
background contribution (see Sect.\,\ref{sec:obs}).  In 4 of the 5 SL
spectra (\object{LkH$\alpha$ 330}, \object{RR Tau}, \object{HD 101412}
and \object{VV Ser}), the features remain after background subtraction
and are concluded to be associated with the source. However, for 1
source, \object{Haro 1-17}, the sky-corrected SL spectrum shows no PAH
features (see Fig.~\ref{fig:allspecs_bgpahs}). The PAH emission seen
towards this source is concluded to be entirely due to background
emission.

A number of the sources were only observed in SH for which the width
of the slit is too small to directly extract the continuum flux
outside the source profile. Here, the background contribution is
estimated from fitting a standard star source profile plus a flat
continuum flux to the measured source profile. This background
subtraction is more difficult, since the small slit length of only
$11.2''$ ($5$ pixels) makes the source profile fits less accurate.

\begin{figure}
  \centering
  \includegraphics[width=\columnwidth]{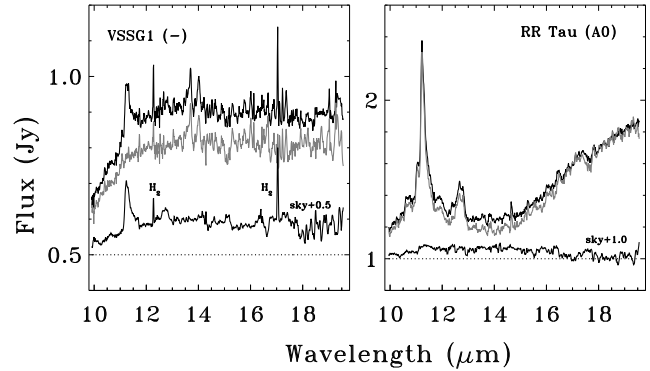}
  \caption{IRS SH spectra for \object{VSSG 1} and \object{RR Tau}. Bottom (black): 
  extracted sky spectrum. Middle (grey): source spectra corrected for background. Top (black): source spectra. The sky spectra are shifted by +0.5 and +1.0 Jy respectively for purpose of clarity.}
  \label{fig:skyspecs}
\end{figure}

For all presented sources, the SH source and sky spectra were
inspected for PAH features.  Two SH background spectra, one with
extended PAH features (\object{VSSG1}) and one without (\object{RR
Tau}) are shown in Fig.~\ref{fig:skyspecs}.
The extracted background spectra for SH show no PAH features for 8 of
the 10 sources in our sample with PAH emission. The two sources with
features in the background spectra are \object{Haro 1-17} and
\object{VSSG1}. In both cases, removing the background emission
entirely removes the 11.2 $\mu$m PAH feature from the source
spectrum (Fig.~\ref{fig:allspecs_bgpahs}). It is concluded that for
these two sources the PAH features are fully due to background
emission.

The environment around one source, the Herbig Ae star \object{VV Ser},
is discussed extensively by Pontoppidan\,et al.\ (2006). Based on IRAC
and MIPS images, it is shown that VV Ser is surrounded by a bright and
extremely large ($\sim$\,$6'$) nebulosity emitting at 8, 24 and 70
$\mu$m, which is not seen in near infrared images. They conclude that
the emission is due to a mix of quantum-heated PAHs and Very Small
Grains (VSGs) present in the low-density nebula. The PAH emission
lines in the spectrum must originate from within 1.8$''$ -- 2.35$''$
or 450 -- 600 AU (half-slit width SL--SH). In their best-fit model
they require a central cavity of 15000 AU, so the extended nebulosity
is not expected to account for the observed PAH features although they note that it
cannot be ruled out that a small clump of PAH material is present
nearer to the star. Here we assume the emission is from PAHs in the
circumstellar disk.

In summary, we conclude that the PAH emission from most of our sources does not originate from extended diffuse fore- or background emission and must instead originate from the observed young stars with disks.

\subsection{Statistics}
\label{sec:stat}
Within our current c2d sample, clear PAH features
are detected in most (5 out of 7) HAeBe stars while only 3 out of 38 T
Tauri stars show features consistent with the presence of PAH
molecules. Interestingly, the PAH detection rate is 100\% for the 4 sources (\object{HD135344}, \object{SR 21N}, \object{LkH$\alpha$ 330} and \object{T Cha}) with SEDs characteristic of cold disks, i.e., sources with SEDs that lack excess emission in the 3--13 $\mu$m region, indicative of an inner hole in the dust disk (Brown et al., in prep). These 4 sources are also the 4 lowest mass sources (spectral type F4 -- G8) with PAH detections.

The c2d sample of PAH detections is biased towards sources with
either a strong $11.2$\,$\mu$m feature or multiple PAH features
from SH and SL observations.  This excludes potential sources with
weak PAH features for which only SH observations are available now: since we
cannot assign the origin of this $11.2$\,$\mu$m feature to either
PAH or crystalline forsterite, these sources are, for now, excluded.
This includes 17 sources with tentative $11.2$\,$\mu$m detections: 14 T Tauri stars, 1 Herbig Ae star and 2 unclassified sources. The 14 T Tauri stars are \object{DoAr 24E}, \object{EC 82}, \object{GW Lup}, \object{GY 23}, \object{HT Lup}, \object{Krautter's Star}, \object{RU Lup}, \object{SX Cha}, \object{SY Cha}, \object{SZ 73}, \object{VW Cha}, \object{VZ
Cha}, \object{V710 Tau} (binary) and \object{WX Cha}.  Future SL data will be able to confirm or
dismiss the presence of PAHs in these sources.  

Finally, in the current sample there are 28 sources with no clear $11.2$\,$\mu$m PAH feature, consisting of 1 Herbig Ae star, 21 T Tauri stars and 6 sources with no known spectral type.
From our present sample, the lower limit to the detection rate of PAH emission features toward T\,Tauri stars is about $8$\,\%. This detection rate goes up to $45\,$\% if tentative detections are included. PAH emission is only detected toward the $3$ more massive T\,Tauri stars of 
spectral type G of our sample, which consists mostly of K--M type stars. 

The rather small fraction ($8$\,\%) of low mass stars in our
c2d sample with PAH emission features, based on the 11.2\,$\mu$m PAH
feature, contrasts with the large fraction \citep[$57$\,\% 
according to][]{ack04} for intermediate mass stars detected
with ISO based on the 6.2\,$\mu$m PAH feature. If only the 11.2
\,$\mu$m features are considered, their detection rate 
drops to 48\,\% based on their Table~3. However, this potentially includes 
sources with only crystalline silicate emission.
\begin{figure}
  \centering
  \includegraphics[width=\columnwidth]{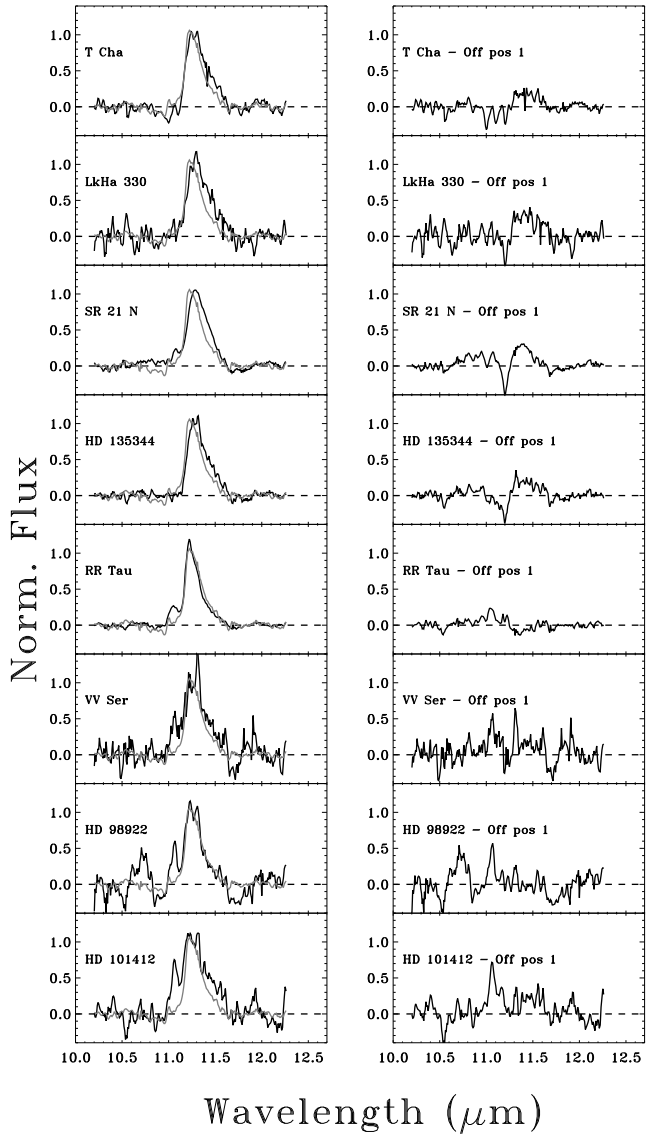}
  \caption{{\bf Left panel:} Blow-up of the SH spectra around the
  continuum-subtracted 11.2 $\mu$m PAH feature, normalized to the
  fitted peak flux (black line). Overplotted in light grey for comparison is the SH
  spectrum of the off-position.  {\bf Right panel:} Plot
  of the difference between the source  and the off-position
  spectra.}
  \label{fig:11p3_overview}
\end{figure}
\begin{figure}
  \centering
  \includegraphics[width=\columnwidth]{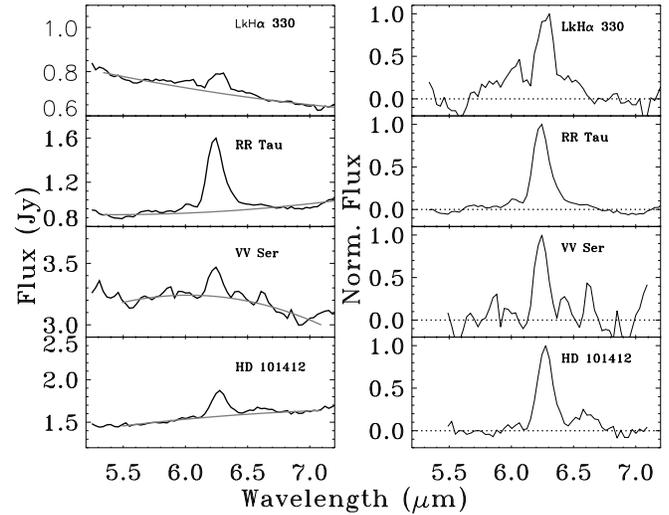}
  \caption{{\bf Left:} Blow-up of the Spitzer-IRS low resolution spectra around the $6.2$\,$\mu$m PAH feature. A simple fit of the continuum flux is plotted with a dotted line. {\bf Right:} Continuum subtracted spectra, normalised to the peak flux of the PAH feature.}
  \label{fig:pah6_specfit}
\end{figure}
\begin{table*}
\centering
\caption{Line fluxes and feature/continuum ratio of PAH features in W m$^{-2}$.}
\label{tbl:lineflux}
\begin{tabular}{lccccccc}
\hline
\hline
\noalign{\smallskip}
  &  \multicolumn{4}{c}{Line flux} & \multicolumn{2}{c}{feature/continuum} \\ 
Name  & \begin{tabular}{c} 6.2\,$\mu$m \\ (SL) \end{tabular} &  \begin{tabular}{c} 11.2\,$\mu$m \\ (SL)\end{tabular} & \begin{tabular}{c} 11.2\,$\mu$m \\ (SH)\end{tabular} & \begin{tabular}{c} 12.8\,$\mu$m \\ (SH)\end{tabular} & \begin{tabular}{c} 6.2\,$\mu$m \\ (SL)\end{tabular} & \begin{tabular}{c} 11.2\,$\mu$m \\ (SH)\end{tabular} & \begin{tabular}{c} cryst. \\ sil.$^{\mathrm{c}}$ \end{tabular}   \\ 
\hline
\noalign{\smallskip}
\object{LkH$\alpha$ 330}  & $\pdx{1.7}{-15}$    & $\pdx{7.7}{-16}$   & $\pdx{5.1}{-16}$       &  $\leq$ $\pdx{2.5}{-16}$ & 1.21 & 1.12 & T \\ 
\object{RR-Tau}  & $\pdx{9.6}{-15}$    & $\pdx{5.0}{-15}$   & $\pdx{ 4.5}{-15}$      &  $\pdx{1.3}{-15}$  & 1.71 & 1.57 & Y \\ 
\object{HD 98922} & -$^{\mathrm{a}}$ &  -$^{\mathrm{a}}$ & $\pdx{1.4}{-14}$     &  $\pdx{4.0}{-15}$  & - & 1.06 & Y \\ 
\object{HD 101412} & $\pdx{4.0}{-15}$    &  $\pdx{2.2}{-15}$   & $\pdx{1.5}{-15}$       &  $\pdx{3.0}{-16}$ & 1.19 & 1.08 & Y$^{\mathrm{d}}$ \\
\object{T Cha}  & -$^{\mathrm{a}}$ & -$^{\mathrm{a}}$ & $\pdx{3.3}{-16}$      &  $\pdx{1.1}{-16}$ & - & 1.15 & N \\ 
\object{HD 135344}          &  -$^{\mathrm{a}}$ & -$^{\mathrm{a}}$ & $\pdx{1.2}{-15}$      & $\pdx{1.1}{-16}$  & -  & 1.19 & N \\ 
\object{EM* SR 21 N}    & -$^{\mathrm{a}}$ &  -$^{\mathrm{a}}$ & $\pdx{4.0}{-15}$      & $\pdx{4.1}{-16}$ & - & 1.32 & T \\ 
\object{VV Ser}    & $\pdx{2.6}{-15}$      & $\pdx{3.1}{-15}$  & $\pdx{2.3}{-15}$      & $\pdx{1.2}{-15}$  & 1.07 & 1.07 & Y \\ 
\hline
Off-position 1       & -$^{\mathrm{a}}$ &  -$^{\mathrm{a}}$ & $\pdx{1.4}{-16}$     &  $\pdx{5.5}{-17}$ & - & & - \\ 
\object{VSSG 1} $^{\mathrm{b}}$ & -$^{\mathrm{a}}$ & -$^{\mathrm{a}}$ & $\pdx{8.6}{-16}$       &  $\pdx{2.4}{-16}$ & - & & Y \\ 
%
\object{Haro 1--17} $^{\mathrm{b}}$          & $\pdx{4.5}{-16}$   &  $\leq$ $\pdx{2.5}{-16}$    & $\leq$ $\pdx{2.5}{-16}$       &  $\leq$ $\pdx{2.5}{-16}$  &  & & Y \\ 
\hline
\end{tabular}
\begin{list}{}{}
\item $^{\mathrm{a}}$ no SL spectra available
\item $^{\mathrm{b}}$ PAH feature detected in background spectrum, not associated with the source 
\item $^{\mathrm{c}}$ crystalline silicates detected in either 28--29, or 33--35 $\mu$m, from Kessler-Silacci et\,al.\ 2006, Table 2; ``Y'' if detected, ``N'' if not detected, or ``T'' if the identification is tentative.
\item $^{\mathrm{d}}$ derived in this study
\end{list}
\end{table*}

\section{Analysis of PAH features}
\label{sec:pahfeatures}

\subsection{Overview of detected PAH features}
\label{sec:pahfeaturesoverview}
Table~\ref{tbl:lineflux} summarizes the detected PAH features and
measured line fluxes. 
The spectrum of \object{RR Tau} nicely shows all the main PAH bands 
detectable in the IRS spectral window at $6.2$, $7.7$, $8.6$, $11.2$,
$12.8$ $\mu$m (Fig.~\ref{fig:allspecs}) and even $16.4$\,$\mu$m (not
shown).  In contrast with the other sources, it has clear $7.7$ and
$8.6$\,$\mu$m features, attributed to C-C stretching transitions and
C-H in plane bending transitions. These features are either less
obviously detected above the silicate continuum or simply absent in
the 4 other sources with SL spectra.  \object{LkH$\alpha$ 330} is a
clear example of a spectrum with a $6.2$ $\mu$m C-C stretching feature and a
$11.2$ $\mu$m C-H out-of-plane bending feature but no $7.7$, $8.6$ nor $12.8$ $\mu$m features. The $7.7$ and $8.6$ $\mu$m features are generally found to be well correlated with the
$6.2\,$$\mu$m feature \citep{pee02} and their absence is thus puzzling in our high sensitivity
spectra. 
\citet{spo02} show that silicate absorption can strongly mask the $8.6$\,$\mu$m PAH feature;
in our case the silicate is in emission, however. According to
\citet{ack04} (see their Table~3), four HAeBe stars observed
with ISO similarly show $6.2$ and $11.2$ $\mu$m emission but no
features at $7.7$ or $8.6$ $\mu$m. Two of these sources, \object{HD
163296} and \object{VV Ser}, are also in the c2d sample (the other two
are \object{HD 142666} and \object{HD 144432}). 
This absence is further discussed in Sect.\,\ref{ssec:diskgeom}, in the context of a disk model 
which demonstrates the lower contrast of the 7.7 and 8.6 $\mu$m features with respect to 
the continuum emission from the disk.

\subsection{Line flux determination}
\label{ssec:lineflux}
To determine the strength of the PAH features, the continuum emission
needs to be subtracted. As a simple approximation we
derive a local pseudo-continuum by fitting a 2D polynomial to the
spectrum around the individual PAH features, where the continuum is
selected by hand.  For the $11.2$\,$\mu$m feature, a polynomial is
fitted to the emission at $10.5$--$11.0$ and $11.8$--$12.2$\,$\mu$m.
The continuum emission below the $6.2$\,$\mu$m emission is estimated
between $5.5$ and $7.1$\,$\mu$m.  We do not include the $7.7$ and
$8.6$\,$\mu$m features in Table~\ref{tbl:lineflux} because these do
not appear significantly in our sources with SL data, with the
exception of \object{RR Tau}. The 12.8 $\mu$m line fluxes extracted from SL
are within the uncertainty consistent with those extracted from SH.

The continuum-subtracted features are integrated between fixed
wavelengths, in particular between $6.0$ and $6.6$\,$\mu$m for the
$6.2$\,$\mu$m PAH feature, between $10.9$ and $11.6$\,$\mu$m for
the $11.2$\,$\mu$m feature (see Fig.~\ref{fig:pah11_sh_specfit} in 
the online material) and between $11.8$ and $13.2$\,$\mu$m
for the $12.8$\,$\mu$m feature. The resulting continuum subtracted
features are shown in Figs.\,\ref{fig:11p3_overview}, \ref{fig:pah6_specfit}  and 
\ref{fig:pah12_sh_specfit}. 

The measured line fluxes are summarized in Table~\ref{tbl:lineflux}.
Thanks to the increased sensitivity of Spitzer, our derived line
fluxes for clearly detected PAH features are an order of magnitude 
lower that what was previously possible with ISO. Our weakest 
detected feature is the $11.2$\,$\mu$m PAH feature 
in \object{T Cha} with a line flux of $3.3 \times 10^{-16}$ W m$^{-2}$. 
A mean 3$\sigma$ sensitivity limit of $2.5 \times 10^{-16}$ W m$^{-2}$ 
is derived from the noise determination in the continuum 
adjacent to the PAH features, though this limit varies somewhat 
from source to source, depending on differences in the S/N of the 
reduced spectra and on the presence of residual reduction artifacts 
for a few cases. For a small number of sources the sensitivity limit 
reaches a few $\times\,10^{-17}$ W m$^{-2}$.

Three HAeBe stars in our sample ---\object{HD 135344},
\object{RR Tau} and \object{VV Ser}--- have previously been observed
with ISO, albeit with much lower $S/N$ ratio \citep{ack04}. 
For \object{RR Tau}, our derived line flux for the $6.2$\,$\mu$m feature agrees 
within $\sim5$\,\% with the ISOPHOT-S spectra, while the $11.2$\,$\mu$m feature is
larger by about a factor 2 in the IRS spectrum. For \object{HD 135344}, we derive a slightly
higher $11.2$\,$\mu$m line flux than the ISO upper limit. For \object{VV Ser}, our derived $6.2$\,$\mu$m line flux is about 4 times weaker, whereas our $11.2$\,$\mu$m detection is consistent with the ISO data.

\citet{slo05} have presented Spitzer SL observations for 4 HAeBe stars with spectra showing PAH features but no silicate dust features, among which \object{HD 135344} is included in our sample. They report clear 6.2, ``7.9'', 11.3 and 12.7 $\mu$m features for all of their sources (\object{HD 34282}, \object{HD 135344}, \object{HD 141569}, \object{HD 169142}). For \object{HD135344} the 7.9 $\mu$m feature is weaker and broader and the 8.6 $\mu$m feature is absent. Their derived line flux for the 11.2 $\mu$m PAH feature is consistent with ours within 10\%.

\subsection{Comparison of PAH features}
\label{sec:featcomp}
In a previous study of a diverse sample of interstellar and
circumstellar sources, planetary nebulae, reflection nebulae
H${\mathrm{II}}$ regions and galaxies \citep{die04}, the
$11.2$\,$\mu$m PAH feature of almost all YSO's, non-isolated HAeBe 
stars and H${\mathrm{II}}$ regions was found to have a similar
asymmetric profile with a FWHM of $\sim$\,0.17\,$\mu$m, and a peak
wavelength in the range of 11.2 -- 11.24\,$\mu$m. The single
isolated HAeBe source in their sample, \object{HD\,179218}, shows a broader 
$11.2$\,$\mu$m PAH feature with a peak wavelength of $\sim$\,11.25\,$\mu$m.  
Our peak position lies in the range 11.25 -- 11.32\,$\mu$m $\pm$ 0.03\,$\mu$m, but this
determination is influenced by the uncertainty in the continuum fit,
and the shift to longer wavelengths may be partially explained by the
presence of a 11.3\,$\mu$m feature from crystalline silicates.

Figure~\ref{fig:11p3_overview} shows a comparison of the
continuum-subtracted $11.2$\,$\mu$m features from all SH
spectra. The off-source PAH feature, which is clearly not contaminated
by silicate emission, is included.  Both the source and the off-source
features are normalized to the peak flux. In the right panel of
Fig.~\ref{fig:11p3_overview} the difference between the two features
is shown.

For \object{T Cha}, \object{LkHa 330}, \object{SR 21 N} and \object{HD
135344}, the $11.2$\,$\mu$m feature is broader than, as well as
redshifted with respect to, the off-position feature. Of these,
\object{T Cha} and \object{HD 135344} show no evidence for crystalline
silicates at 28-29 and 33.6\,$\mu$m (Kessler-Silacci et al.\ 2006),
while \object{LkH$\alpha$ 330} and \object{SR 21 N} show tentative
crystalline features. Thus, the presence of a $11.3$\,$\mu$m
crystalline silicate feature cannot readily explain the broadening of
the measured feature for these sources.  Such a broad shape has been
seen before in the planetary nebulae \object{IRAS 17047-5650} and
\object{IRAS 21282+5050} \citep{hon01}.  \citet{pec02} proposed
anharmonicity as an explanation of the broadening and used a PAH
emission model to fit the 11.2 $\mu$m feature of IRAS 21282+5050
with a combination of the fundamental ($v = 1 \rightarrow 0$) and hot
bands ($v = 2 \rightarrow 1$ and $v = 3 \rightarrow 2$) of the
transition. These hot bands would point to very hot PAHs being
present, presumably in the innermost part of the disk where the
radiation field is strongest.

For the HAeBe sources \object{RR Tau}, \object{VV Ser}, \object{HD
98922} and \object{HD 101412}, the peak wavelength of the $11.2$ PAH
feature is very similar to that of the off-position feature. For
\object{RR Tau}, the PAH feature compares very well with the
off-source feature, both in shape and peak position, showing little to
no residual after subtraction. For the other 3 sources, subtracting
the off-position feature leaves several residual features, hinting at
the presence of crystalline features, which are also seen at longer
wavelengths.

\section{PAH emission from disks}
\label{sec:pahindisks}

\subsection{Disk model}
\label{ssec:diskmodel}
The strength of the PAH emission features is known to depend on the
strength of the UV and optical radiation field, but in disks several
additional parameters can affect the appearance of the PAH
features. Here we address the question as to how the PAH
features are affected by the spectrum of the central source, the PAH
abundance and the flaring geometry of the disk. A related question is
why no PAH features are seen in at least half of our sample of T Tauri
disks, in contrast to the findings for HAeBe stars.

We use the 3-dimensional Monte Carlo radiative transfer code RADMC
\citep{dul04}, for which a module to treat the emission from quantum-heated 
PAH molecules and Very Small Grains has been included. This module will 
be described in detail in Dullemond et al.~(in prep.), but a rough description 
has been given by Pontoppidan et al.~(2006). A template model is set up using
the following model parameters. A Kurucz model spectrum is taken for
the central star with $T_{\mathrm{eff}}$ = 10000 K, but the stellar
parameters (luminosity, stellar radius) were chosen from evolutionary
tracks by \citet{siess00} for an age of 3 Myr. The disk is modeled
with $M_{\mathrm{disk}} = 1\times10^{-2}M_{\odot}$, an inner radius
set by a dust evaporation temperature of 1300~K, and an outer radius of
300 AU. The disk is flaring, with the vertical pressure scale height
(in units of radius) at the inner rim $H_{p}/R_{\mathrm{in}}$ = 0.02 and at
the outer edge $H_{p}/R_{\mathrm{out}}$ = 0.14. The disk is modeled to be close
to face-on ($i=5.7^{\circ}$), to maximize the
strength of the PAH features. The spectra are scaled to a distance to
the observer of 150 pc.

The PAH emission is calculated for an equal mix of neutral
and singly ionized C$_{100}$H$_{24}$ molecules, adopting the
\citet{dra01} PAH emission model, using the ``thermal continuous''
approximation. Multi-photon events are included for the PAH
excitation, following the method outlined by
\citet{sie02}. Enhancement factors for the integrated cross sections
of the 6.2, 7.7 and 8.6 $\mu$m bands ($E_{6.2}=3$, $E_{7.7}=2$,
$E_{8.6}=2$) as suggested by \citet{li01} are taken into account, as
also implemented in H04. The actual ionization state of
PAH, which varies as a function of location in the disk, can
affect the 7.7 and 8.6 $\mu$m features relative to 11.2 and 3.3
$\mu$m but this will be explored further in future models
which include a full ionization balance of multiple PAH species in
disks (Visser et al.\ in prep). We include the opacities from
\citet{mat05a} for near-infrared wavelengths. Several model
tests were performed, which are presented in Appendix
\ref{sec:habcomp}, including a comparison with H04.

PAH destruction is expected to occur in a strong UV radiation field,
when the PAH molecules, through multi-photon events, absorb more than
21 eV in an interval shorter than their cooling timescale
\citep{guh89}.  If the PAH destruction happens on a shorter timescale
than the lifetime of the disk, this can have an effect on the PAH
abundance in, and expected PAH emission from, the inner region of the
disk. This effect will be stronger for smaller ($N_{\mathrm{c}}<50$)
PAHs, and depends on the assumed temperature at which the PAHs
dissociate. Model calculations by Visser et al.\ (in prep.) show
that for PAHs with $N_{\mathrm{c}} = 100$ the lifetime is larger than
the lifetime of the disk, and therefore the PAHs are kept at constant
abundance throughout the entire the disk.
Tests with removing PAHs inwards from a destruction radius 
$R_{\mathrm{PAH, in}}$ (by setting their abundance to zero) show that 
the results do not depend sensitively on the choice of $R_{\mathrm{PAH,in}}$
when it is of order 1 AU.

In our template model, we use the PAH abundance of H04, who
adopt a fraction of 23\,\% of available carbon being locked up in
C$_{100}$H$_{24}$ PAHs, corresponding to a carbon abundance with
respect to hydrogen of $5 \times 10^{-5}$.  For a single PAH species
with $N_{\mathrm{C}} = 100$ (C$_{100}$H$_{24}$) this leads to an
abundance of $5 \times 10^{-7}$. In our model the abundance is set as
a fraction of dust mass.  Assuming C$_{100}$H$_{24}$ (molecular mass:
$2.04 \times 10^{-21}$g) and a dust-to-gas ratio of 1 to 100, the PAH
abundance of $5 \times 10^{-7}$ with respect to hydrogen corresponds
to a mass fraction of
0.061 gram of PAH per gram of dust, of which 50\% ionized and 50\% neutral.
Such an abundance is at the high end of that inferred for general
interstellar clouds \citep{ces00,hab01}.
Indeed, the parameters chosen in our model and that of H04
maximize the PAH emission.

\subsection{Dependence on spectral type}
\label{ssec:stellaruv}
\begin{figure}
  \centering
  \includegraphics[width=\columnwidth]{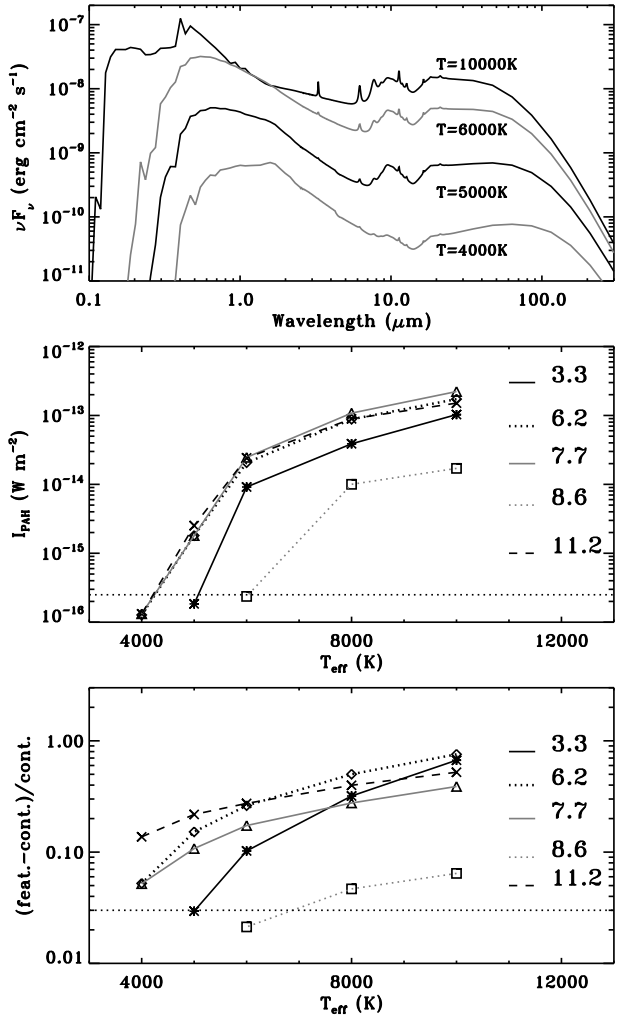}
  \caption{{\bf Top panel:} Model SEDs of a star+disk with PAHs, for 
different spectral types of the star, with $T_{\mathrm{eff}}$ = 10000, 
6000, 5000 and 4000 K at a distance of 150 pc. {\bf Middle panel:} 
PAH line flux of the various features for the corresponding 
models, including also $T_{\mathrm{eff}}$ = 8000 K. {\bf Bottom panel:} 
Feature to continuum ratio of the PAH features.
 The dotted line indicates our $3\sigma$ observational limits.}
  \label{fig:pahsttype}
\end{figure}

\begin{figure}
  \centering
  \includegraphics[width=\columnwidth]{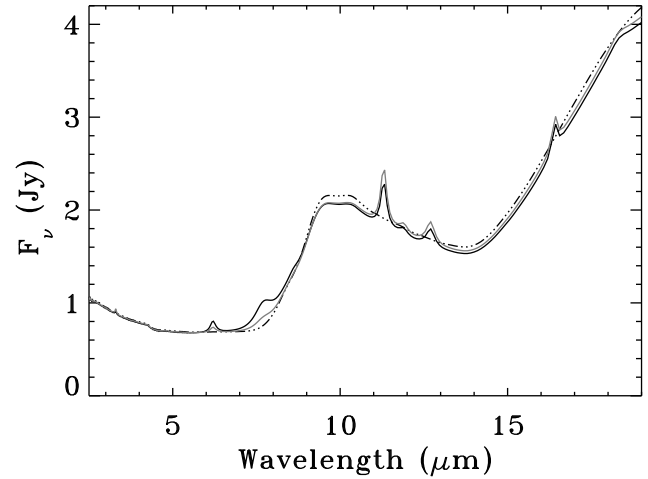}
  \caption{Blowup of the standard model spectrum of a
  star+disk with PAHs, for $T_{\mathrm{eff}}$ = 6000 K with template
  parameters (dark solid line), without PAHs (dash-dotted line), with
  only neutral C$_{100}$H$_{24}$ PAH (light solid line). Note that the 11.2
  and 6.2 $\mu$m features are clearly visible on top of the continuum,
  whereas the 7.7 and 8.6 $\mu$m features are masked by the rising 10
  $\mu$m silicate feature.}
  \label{fig:pahsttype_zoom}
\end{figure}

Since UV and optical radiation incident on the disk surface provides
the main PAH excitation mechanism, the PAH features must depend on the
spectral type of the illuminating star.
To address this influence our
template disk model is calculated for a range of Kurucz stellar
spectra, with $T_{\mathrm{eff}}$ = 10000, 8000, 6000, 5000 and 4000
K, corresponding to spectral types A0, A6, G0, K2 and K7, respectively
\citep{gra94}. Stellar parameters are again taken from
evolutionary tracks \citep{siess00} for an age of 3 Myr. The disk
parameters were modified slightly in 1 iteration to ensure hydrostatic
equilibrium throughout the disk, except for the inner rim. The model
SEDs for the different central stars and absolute continuum subtracted
fluxes for the PAH features are presented in
Fig.~\ref{fig:pahsttype}. A blowup of the standard model spectrum is
shown in Fig.~\ref{fig:pahsttype_zoom}.

Since the absolute strength of the PAH features scales 
foremost with the total radiation that is absorbed by the 
PAHs, it also depends on disk parameters that are 
unrelated to the PAHs themselves. To evaluate the role of the disk 
continuum, we present feature/continuum ratios
in Fig.~\ref{fig:pahsttype} as well.

It is seen that with decreasing $T_{\mathrm{eff}}$ both the line flux
as well as the feature/continuum ratio decrease for the 3.3, 6.2, 7.7,
8.6 and 11.2\,$\mu$m features (Fig.~\ref{fig:pahsttype}).  The rate
of decrease in line flux is similar for all features, as is that of the
feature/continuum ratio except
for the 3.3 $\mu$m feature. The latter feature disappears more
rapidly with decreasing $T_{\mathrm{eff}}$ because the continuum
emission at 3.3 $\mu$m decreases at a slower rate compared to the
longer wavelengths. For all $T_{\mathrm{eff}}$ considered,
the 8.6\,$\mu$m feature has the lowest feature to continuum ratio,
from $\lesssim$ 7 to $\lesssim$ 3\% , which can be clearly
seen in Fig.~\ref{fig:pahsttype_zoom}. 
When the PAH molecules are assumed to be 100\% neutral
C$_{100}$H$_{24}$, the feature/continuum ratio of the 6.2, 7.7 and 8.6
$\mu$m bands decreases by a factor $\sim$\,2--3, while that of the 3.3 and 11.2
$\mu$m features increases by a factor $\sim$\,1.5--3 (Fig.~\ref{fig:pahsttype_zoom}).

These models show that the $6.2$ and $11.2$ $\mu$m features
are the most suitable tracers for the presence of PAHs in disks around
late type T Tauri stars owing to their relatively high
feature/continuum ratio, and this qualitatively agrees with the fact
that these two features are always present in our T Tauri stars with
PAH detection.  Quantitatively, Fig.~\ref{fig:pahsttype} shows that,
for the current model assumptions, the PAH features become difficult
to observe for $T_{\mathrm{eff}} < 4200$ K (later than K6)
even with a relatively high PAH abundance of $\sim$\,6\% of the
dust mass in the entire disk and large flaring angles.
\begin{figure}
  \centering
  \includegraphics[width=\columnwidth]{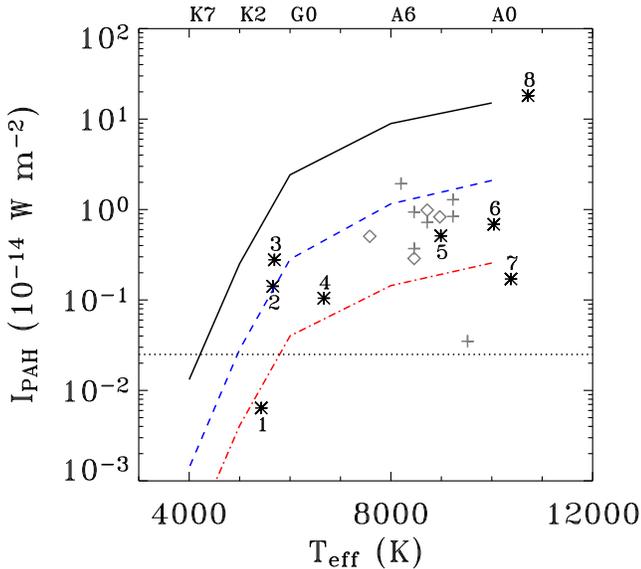}
  \caption{Strength of 11.2 $\mu$m PAH feature $I_{\mathrm{PAH}}$ (scaled to
  a distance of 150 pc) versus $T_{\mathrm{eff}}$. Black 
  solid line represents our template model, based on Fig.~\ref{fig:pahsttype}; 
  dashed and dash-dotted lines represent our models with a 10x and 100x lower PAH abundance respectively.
  Black `*' symbols are for   c2d sources. Grey diamonds are ISO upper limits of the 
  $11.2$\,$\mu$m feature strength for Herbig Ae sources, grey `+' symbols are ISO detections. 
c2d sources are labelled as follows. 1: 
  \object{T\,Cha}; 2: \object{LkH$\alpha$\,330}; 3: \object{SR\,21 N}; 
  4: \object{HD\,135344}; 5: \object{RR\,Tau}; 6: \object{VV\,Ser}; 
  7: \object{HD\,101412}; 8: \object{HD\,98922}. The dotted grey line indicates 
  our typical 3$\sigma$ sensitivity limit of $2.5 \times 10^{-16}$ W m$^{-2}$ for 
  sources at $d = 150$ pc.}
  \label{fig:pah11_str_temp}
\end{figure}
\begin{figure}
  \centering
  \includegraphics[width=\columnwidth]{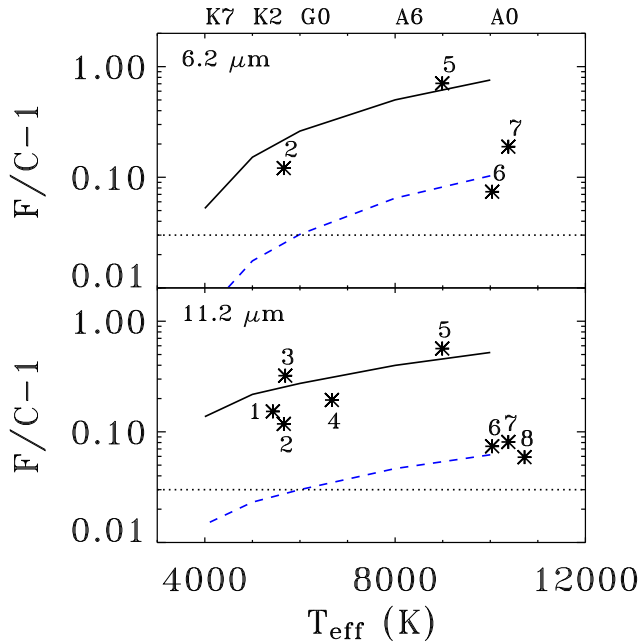}
  \caption{Feature / continuum ratio for the 
   6.2 and 11.2 $\mu$m PAH 
  features; solid line represents our template model, based on 
  Fig.~\ref{fig:pahsttype}; dashed line represents our model with a 
  10x lower PAH abundance.}
  \label{fig:f_c}
\end{figure}

The observed $11.2$\,$\mu$m line fluxes have all been rescaled to a
distance of $d=150$ pc, and are plotted in Fig.~\ref{fig:pah11_str_temp}
together with the models. Also included is our typical 3$\sigma$ limit
for a source at 150 pc. For comparison we added a small subsample of
12 Herbig Ae stars, selected from \citet{ack04}, and rescaled to 150
pc.  Stars with unknown distance and/or poorly known spectral type
were excluded.  The observed feature/continuum ratios for the two
strongest features, 6.2 and 11.2\,$\mu$m, are compared with models
in Fig.~\ref{fig:f_c}.

Figure~\ref{fig:pah11_str_temp} shows that our template model
(solid line) predicts a larger $I_{\mathrm{PAH}}$ than observed,
with the exception of one source (\object{HD~98922}) for which the
distance is uncertain.  A lower observed 11.2 $\mu$m flux
can be caused by a number of conditions such as a lower PAH abundance,
a smaller flaring angle or a disk orientation close to edge-on
(see below).  Included in Fig.~\ref{fig:pah11_str_temp} are two
additional model runs, where the PAH abundance has been lowered by a
factor of 10 (dashed line) and 100 (dash-dotted line). Most of the
observations fall within the predictions of these two models. Only the
model with a 100$\times$ lower PAH abundance predicts a 11.2 $\mu$m
feature strength for G-type stars below our Spitzer detection limit at
$d=150$ pc.  One source, \object{T Cha}, is detected despite the fact
it falls below this formal detection limit because it has a distance
of $d=66$ pc.

Figure~\ref{fig:f_c} indicates that the template model fits the
observed 11.2\ $\mu$m feature/continuum ratios for the 4 low mass
sources, even though it overpredicts $I_{\mathrm{PAH}}$. The model
with 10$\times$ lower PAH abundance fits the feature/continuum ratio
of three of our HAeBe stars. However both models with 10$\times$ and
100$\times$ lower (below plot limit) PAH abundance predict a
feature/continuum ratio below our Spitzer detection limit for spectral
types below G0. This suggests that the small number of T Tauri
detections (3 out of 38) are presumably outliers, with abnormally high
feature/continuum PAH features.  This discrepancy will be further
discussed in Sect.~\ref{ssec:diskgeom} and is most likely due to an
abnormally low continuum in these sources.

\subsection{Additional UV radiation and relation with H$\alpha$}
\label{ssec:adduv}
The models presented in Sect.~\ref{ssec:stellaruv} include only the
stellar radiation, but not any additional sources of UV.  Excess UV
radiation compared with the stellar photosphere has been observed from
at least some T Tauri stars (e.g., Herbig \& Goodrich 1986, Costa et
al.\ 2000, Bergin et al.\ 2003). From a constructed composite FUV
spectrum, derived from two K7 sources representative of low-mass T
Tauri stars, \citet[ and references therein]{ber03} find an overall
FUV continuum flux at $r=100$ AU on the order of a few hundred times
the Habing field. The effect of this additional UV is to
shift absolute feature strengths to those appropriate for higher
$T_{\mathrm{eff}}$, close to $T_{\mathrm{eff}} \sim 10000$~K.  If a
large fraction of T Tauri stars would have such additional UV
radiation this leads to the question why not more sources were
detected, since the feature strength should be more than sufficient if
PAHs are present at our template abundance.
This would further strengthen the evidence for low PAH abundances
in a large fraction of T Tauri disks.

The origin of excess UV radiation from T Tauri stars is not
fully clear, but is usually thought to originate from the shock in the
magnetospheric accretion column associated with material falling from
the inner disk onto the stellar surface.  One indication of accretion
activity is the strength of the H$\alpha$ emission line, measured in
H$\alpha$ equivalent width (EW) \citep{cab90}. Literature values of
H$\alpha$ EW measurements are listed in Table~\ref{tbl:obssum}.  In
our sample, \object{LkH$\alpha$ 330} is a classical T Tauri star
(H$\alpha$ EW $\geq 10~\AA$), while \object{T Cha} should be
classified as a weak line T Tauri star (H$\alpha$ EW $< 10~\AA$).  We
find no correlation between the $11.2$\,$\mu$m PAH feature strength
and H$\alpha$ EW.  However, single measurements of H$\alpha$ EW may
not necessarily be a reliable tracer of the average accretion
luminosity. As an example, for \object{T Cha} the quoted H$\alpha$ EW
of 2~$\AA$ by \citet{alc93} is indicative of a low accretion rate,
while \citet{syl96} have later classified it as a YY Orionis star with
strong UV continuum emission and substantial variability on short
timescales. Another problem is that not all of this UV
radiation from the accretion shock may reach the disk, but can be
absorbed by the overlying accretion column and/or any bi-polar outflow
material \citep{ale05}.

An alternative interpretation of the enhanced UV radiation is stellar
activity, especially at later evolutionary stages when the accretion
becomes less important. This mechanism has been used by \citet{kam04a}
to model the UV field around young G-type stars.

\subsection{PAH abundance}
The discussion in Sect.~\ref{ssec:diskmodel} has illustrated
some of the effects of the PAH abundance on the PAH features.
As seen in Figs.~\ref{fig:pah11_str_temp} and \ref{fig:f_c}, the PAH
features decrease in line flux and feature to continuum ratio with
decreasing PAH abundance, with a similar trend for all the PAH
features. Decreasing the PAH abundance also affects the overall SED, increasing
the continuum radiation at wavelengths longer than 2\,$\mu$m and at
UV wavelengths around 0.2\,$\mu$m.

For sources with detected 11.2 $\mu$m PAH feature, the line
strengths indicate PAH abundances between 10 and 100$\times$ lower
than the template model abundance, i.e., between 0.06 and 0.6 \% of
dust mass. With respect to hydrogen, this translates to PAH abundances
between $5\times 10^{-9}$ and $5\times 10^{-8}$. Geometry affects
these numbers by at most a factor of a few (see 
Sect.~\ref{ssec:diskgeom}). Thus, the absence of PAH features in the
majority of our T Tauri sources may be partly caused by a lower PAH
abundance than found in molecular clouds.  However, this does not mean
that PAHs are absent; even at these lower abundances, PAHs can still
have an important influence in terms of UV opacities and heating rates
in the surface layer of the disk (Jonkheid et al., submitted). For comparison,
the PAH abundance inferred for the HD 141569 transitional disk is
0.00035\% compared to dust \citep{li03} or $1.5\times 10^{-10}$ with
respect to hydrogen \citep{jon06}. 

\subsection{Disk geometry}
\label{ssec:diskgeom}
\begin{figure}
  \centering
  \includegraphics[width=0.9\columnwidth]{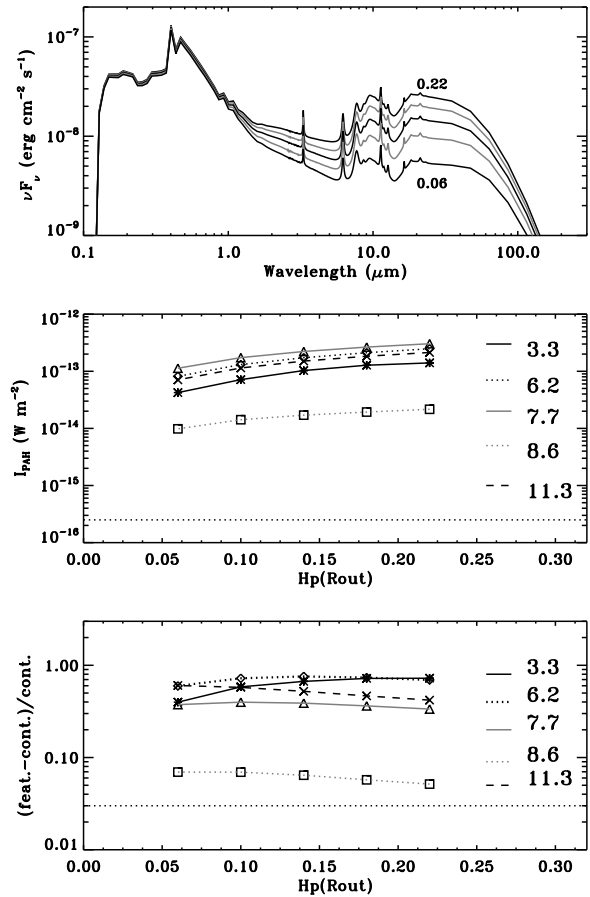}
  \caption{{\bf Top panel:} Model SEDs of a star+disk with PAH, with 
scale height of outer disk $H/R_{\mathrm{out}} = 0.06$, 0.1, 0.14, 0.18 
and 0.22 at a distance of 150 pc. {\bf Middle panel:} PAH line fluxes 
for the corresponding models. 
{\bf Bottom panel:} Feature to continuum ratio 
of the various features.
 The dotted line indicates our observational limits.}
 \label{fig:pahflarecomp}
\end{figure}

Finally, the effect of the disk geometry on the PAH features is
addressed. The template disk and PAH model is used, again with $T_{\rm
eff}=10000$~K. 
The pressure scale height of the disk at the outer radius is varied
between $H_{p}/R_{\mathrm{out}} = 0.06$ and 0.22, while the pressure
scale height of the inner rim is kept constant at
$H_{p}/R_{\mathrm{in}} = 0.02$. This illustrates the effect on the PAH
features of a flaring disk versus a flatter disk. SEDs for varying
values of the scale height of the outer disk are shown in
Fig.~\ref{fig:pahflarecomp}.
The line flux of the PAH features is found to slowly decrease
as the disk becomes less flared by lowering the outer disk scale
height from $H/R_{\mathrm{out}}$ = 0.22 to 0.06. The feature over
continuum ratio is found to increase slightly with decreasing
$H/R_{\mathrm{out}}$ for most features, with the exception of the 3.3
and 6.2 $\mu$m features, where the feature/continuum ratio decreases
for $H/R_{\mathrm{out}}<0.1$. This trend is caused by a more rapid
decrease of the silicate continuum flux compared to the PAH peak
flux. All the main PAH features can be distinguished for the flaring
disk, although the 8.6 $\mu$m feature has a low feature to continuum
ratio of $\sim$\,3--10\%.  

To compare this scenario to our PAH detections, a measure for the
flaring index of the disk is taken to be the shape of the SED,
obtained by dividing the observed $\nu F_{\nu}$ at 35 $\mu$m over 13
$\mu$m \citep{kes06}. This ratio is compared to the measured
11.2\,$\mu$m line flux.  No clear evidence for a correlation between
the 11.2\,$\mu$m line strength and a rising/non-rising SED is
found. Together with the lack of strong changes
Fig.~\ref{fig:pahflarecomp}, this suggests that disk flaring
alone is not sufficient to explain the PAH non-detections in our
sample.

A second geometry effect, disk thickness, can have a direct 
effect on the PAH feature strength $I_{\mathrm{PAH}}$. A geometrically 
thicker disk captures a larger portion of the stellar radiation field than 
a flatter disk, and hence the response in {\em both} the dust continuum 
{\em and} the PAH features is stronger. Observations of spectral energy 
distributions of disks around pre-main-sequence stars have indicated 
that disk geometries vary strongly from source to source \citep{mee01}, 
and that this factor can therefore potentially affect the spectrum.
Test model runs show that increasing the disk geometrical thickness by
a factor 2, by increasing the pressure scale height $H_p$ throughout
the entire disk, introduces an increase in $I_{\mathrm{PAH}}$ of
$\sim$\,1.5--2, while the ratio of PAH feature over 
continuum ratio decreases by a factor $\sim$\,1.4--1.8. The disk would have to be a factor $\sim$\,60--80 thinner for the PAH features to have been undetectable for $T_{\mathrm{eff}} < 6000$ K (G0), for the template abundance.

A third geometry effect, inclination, affects both $I_{\mathrm{PAH}}$ and the feature over continuum ratio. Our template models are run for face-on disks ($i = 5^{\circ}$). When the template model is observed at angles up to $i \sim 60^{\circ}$, $I_{\mathrm{PAH}}$ remains relatively unchanged ($<$ 3\%) and the feature over continuum increases by about 10\%. For an almost edge-on disk ($i = 80^{\circ}$), $I_{\mathrm{PAH}}$ has decreased by $\sim$\,40\%, while the feature over continuum is a factor $\sim$\,4 larger. Inclination may be an explanation for some of the sources with abnormally large F/C ratio, compared with their feature strength,  e.g. \object{RR Tau}.

A fourth geometry effect which could explain an unusually high
feature/continuum ratio is the introduction of a large scale gap in
the inner dust disk, for example as the result of dust coagulation or
possibly through clearing by one or more planet(s), although the
latter would require extreme circumstances. This lowers the 3--13
$\mu$m continuum flux while keeping the PAH feature strength the
same.  The motivation for this explanation comes from the SED of the
small fraction of T Tauri stars in our sample that are detected, all
of which have a higher 11.2 $\mu$m feature/continuum ratio than would
be consistent with their 11.2 $\mu$m fluxes within the standard
model. The SEDs of these 4 detected sources show remarkably low
emission at 10 $\mu$m followed by a steep rise towards 20 $\mu$m,
suggestive of a gap opening up in the dust disk (Brown et al., in
prep.).  Examples of similar SEDs include those of DM Tau, GM Aur and
CoKu Tau/4 \citep[e.g.][]{dal05}, for which 
the continuum flux around 11 $\mu$m is also depressed by more than an
order of magnitude compared with that of a standard T Tauri star in
Taurus.
 
\subsection{Model summary}
In summary, all three effects of reducing the stellar temperature (and
thus the amount of optical and UV radiation), reducing the PAH
abundance, and decreasing the flaring angle or thickness of the disk
are shown to decrease the absolute strength of the PAH features, as
expected.  The feature over continuum ratio trends are similar for the
6.2, 7.7, 8.6 and 11.2\,$\mu$m PAH features. 
That of the 8.6\,$\mu$m is the lowest in all models,
with typical values of $<$ 6--10\% even for models which maximize the
features, which can qualitatively explain the observed lack of this
feature.

The 11.2 $\mu$m feature remains one of the strongest features in all
models explored and this is qualitatively consistent with our
observations. For the template model parameters which maximize the PAH
features, PAHs are observable within the mean sensitivity limit of our
observations for $T_{\mathrm{eff}} \ge 4200$ K (K6). 
 For a 10$\times$ lower PAH abundance, the features
are still above our sensitivity limit for $T_{\mathrm{eff}} \ge 4900$ K (K2), 
but now the feature/continuum
ratio becomes the limiting factor, with values $\lesssim$ 5\% for
$T_{\mathrm{eff}} \le 6000$ K (G0).  This partly explains the
absence of any PAH features in the spectra of more than half of the
sources in the sample considered, which consists mostly of late K and
M type stars.

\section{Conclusions and future work}
\label{sec:conclusions}
Spitzer-IRS spectra were obtained for a set of 54 pre-main sequence
stars with disks, including 38 T Tauri stars and 7 Herbig Ae
stars. The observations are an order of magnitude more sensitive than
those used in previous surveys of PAHs in disks. We detect PAH
features in at least 3 T Tauri stars, with an additional 14 tentative
detections to be confirmed, resulting in a lower limit to the PAH
detection rate in T Tauri stars of 8\,\%.  Spitzer SL observations are
needed to confirm the presence of the PAH features for sources where
we currently only see the $11.2$\,$\mu$m feature in the SH
spectrum. All 4 sources that show hints for inner holes in
their dust disk also a show clear 11.2\,$\mu$m PAH feature.

The lowest mass source with PAH emission in our sample is \object{T
Cha} with spectral type G8.  The derived 11.2 $\mu$m line
intensities are between a few $\times 10^{-15}$ and $3.3 \times
10^{-16}$ W m$^{-2}$, which is typically an order of magnitude lower
than what was observed for HAeBe stars with ISO.

Radiative transfer modeling of disks coupled with PAH emission models
shows that for stars of late spectral type, the $11.2$ $\mu$m
feature is expected to be the best tracer of the presence of PAHs. The
models also show that the 7.7 and 8.6 $\mu$m PAH features are most
affected by veiling by the continuum due to strongly rising silicate
emission, resulting in low feature over continuum ratios.

For the small number of T Tauri sources detected as well as for the
Herbig stars, the measured PAH line fluxes and
feature/continuum ratios are lower than those found from our template
disk model which maximizes the PAH emission. Variations of the model
parameters indicate that the most likely explanation is lower PAH
abundances by factors of 10--100, with geometry affecting these
conclusions at the level of a factor of a few. The high
feature/continuum ratios for the detected T Tauri stars are 
due to their abnormally low continuum at 11 $\mu$m caused by
the dust holes in their inner disks.

The template model predictions indicate that the 11.2 $\mu$m feature
strength becomes undetectable at our sensitivity limit when $T_{\rm
eff}<4200$~K (K6). This likely explains the absence of PAH features
for the majority of sources in our sample which have spectral types
later than K0 ($\sim$\,70\%). If a large fraction of these sources would
have excess UV radiation at the level detected for some K7 stars, then
the absence of PAH features implies a lower PAH abundance by at least
an order of magnitude. The same conclusion holds for the 11 out of 38
sources with spectral types earlier than K6 which are not detected.
Geometry affects these conclusions by a factor of a few.  Thus, the
lack of PAH detections does not mean that PAHs are absent in these
disks; even at lower abundances they do need to be considered since they
can affect disk structure, chemistry and gas heating.

In a follow-up paper, the combination of 5-35\,$\mu$m spectra for
the entire sample with IRAC and MIPS photometry will be used to
improve the uncertainty in the PAH detection rate and to compare the
PAH sources in terms of their disk SEDs and interpret this with the
disk modeling.

Ground-based spectra using instruments like VLT-ISAAC and VLT-VISIR on
8-m class telescopes will be able to both search for presence of the
$3.3$\,$\mu$m PAH feature as well as better characterize the shape
of the 8.6 and 11.2 $\mu$m PAH feature through higher spectral
resolution. Through higher spatial resolution, they will also
permit us to put a stronger constraint on the spatial extent of the
PAH features from T Tauri disks.

\begin{acknowledgements}
Support for this work, part of the Spitzer Legacy Science Program, was
provided by NASA through contracts 1224608, 1230779 and 1256316 issued
by the Jet Propulsion Laboratory, California Institute of Technology,
under NASA contract 1407. Astrochemistry in Leiden is supported by a
NWO Spinoza grant and a NOVA grant, and by the European Research
Training Network "The Origin of Planetary Systems" (PLANETS, contract
number HPRN-CT-2002-00308). B.\ Mer\'{\i}n acknowledges
funding from the ``Fundaci\'on Ram\'on Areces (Spain). The authors
wish to thank Louis Allamandola and Andrew Mattioda for discussions
and new opacities and Emilie Habart for comparisons with her models.
\end{acknowledgements}

\bibliographystyle{aa}
\bibliography{4830}

\appendix
\section{Model tests and comparison with Habart et al.~2004}
\label{sec:habcomp}
There are various differences in the modeling approach
between the \citet{hab04} paper and our models, which are summarized
in Table~\ref{tbl:modelcomp}. First, the H04 model is a 1+1-D model,
meaning that the model consists of a series of vertical 1-D models at
different radii, combined to make a full disk structure. This method
follows the irradiation-angle philosophy used in many other models as
well, e.g.\ \citet{dal98}, \citet{bel97}, \citet{dul02}.  In our model
we use full 3-D radiative transfer based on an axisymmetric disk
density structure. Our model therefore also includes emission from the
dust inner rim, contrary to the approach of H04. Our 3-D approach
allows radiative energy to be exchanged between adjacent radii, which
is not possible in the 1+1D approach.  Consequently the model SEDs
differ somewhat between these two types of models.  Specifically, for
a flaring disk, the radiation will more readily escape in the polar
direction than in the radial direction, because of the larger optical
depth in the latter direction.
When the disk is viewed face-on, the SED will be boosted in the 3-D
approach compared to the 1+1-D approach by up to $\sim$\,40\%.
Moreover, the axisymmetric 3-D models described here allow the
treatment of radiative transfer in disk geometries that are not
flaring, e.g.\ self-shadowed disks (see Section \ref{sec:pahindisks}
for examples).  On the other hand, in our models the vertical
geometric thickness of the disk is only {\em estimated} to be roughly
consistent with hydrostatic equilibrium, whereas the H04 model
includes detailed hydrostatic equilibrium. The results of these two
approaches, however, are nearly identical.

There are also fundamental differences in the opacities and
the treatment of dust grains in the radiative transfer. Most
importantly, H04 thermally decouple the carbon grains from the
silicate grains, while we have both grain types at the same
temperature and mix the opacities into a single opacity. 
Our assumption may be more realistic because silicate and
carbon grains can only have different temperatures at the same
location in the disk if such grains are physically disconnected. Only
a small amount of coagulation is required to get these grains in
physical contact, forcing them to have the same temperature. 
The result is that the silicate feature strength over the continuum is
much stronger in our case than in the case of H04, as can be seen in
their Figure~3 as compared to our Fig.~\ref{fig:pahsttype_zoom}. If
the graphite is decoupled from the silicate, the graphite heats up and
the silicate cools down due to the much lower optical/NIR opacity of
silicate compared with graphite. The superheated graphite emission
therefore fills in the spectrum on both sides of the silicate feature
in the case of thermally decoupled grains.

For stars with lower effective temperature, radiation at
longer wavelengths becomes more inportant. Our models include the new
PAH opacities of \citet{mat05a} at optical and near-infrared
wavelengths, which are higher than those adopted by H04.  In our model
we do not include PAH-destruction in the inner disk, in contrast with
H04. Test models with and without PAH destruction show only small
differences, however.

The biggest differences between H04 and our models are likely
due to the treatment of the stellar radiation.  Our models assume that
the full stellar flux can irradiate the disk, whereas H04 assume that
the disk goes all the way to the stellar surface and occults the lower
half of the star. Hence the H04 models have only half the illuminating
flux as ours.  Moreover, they include the star as a blackbody emitter
with main sequence stellar parameters, while in our model we use a
Kurucz model with pre-main sequence values.

To quantitatively compare the two models, we have plotted our
PAH intensities in the same way as H04, as functions of the integrated
stellar radiation field in the FUV (6--13.6 eV, 912--2050 $\AA$)
wavelength range. The UV flux is usually specified relative to the
average interstellar radiation field from \citet{hab68} integrated
over this wavelength range, which is $1.6 \times 10^{-3}$ erg
cm$^{-2}$ s$^{-1}$ \citep{tie85}. This quantity is most often referred
to as $G_0$ but we refer to it here as $\chi$ to be consistent with
H04. H04 assume a single blackbody for the stellar radiation field for
stars of spectral type B5--G0 and rescale the flux from the stellar
surface $R_*$, to the flux at a distance from the star of 150 AU, as a
reference point in the middle of their disk.

\begin{figure}
  \centering
  \includegraphics[width=\columnwidth]{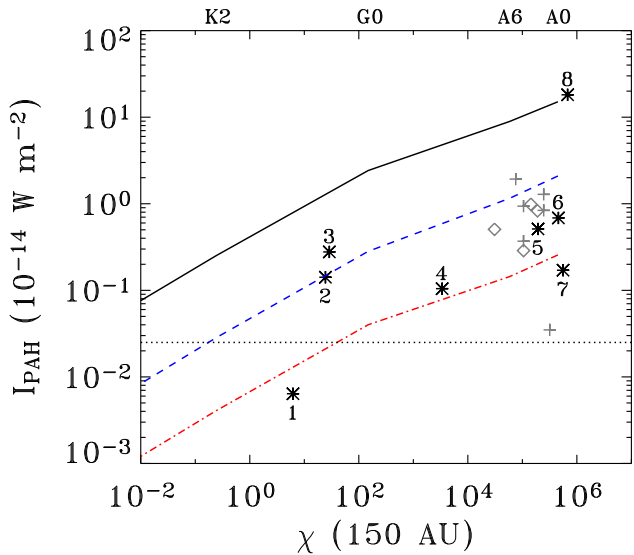}
  \caption{Strength of 11.2 $\mu$m PAH feature $I_{\mathrm{PAH}}$ (scaled to
  a distance of 150 pc) versus $\chi$, the integrated FUV field (6--13.6eV) 
  for a ZAMS star, after \citet[][ their Figure~8]{hab04}. Solid line represents our template model, based on Fig.~\ref{fig:pahsttype}; 
  dashed and dash-dotted lines represent our models with a 10x and 100x lower PAH abundance respectively.
  Black `*' symbols are for   c2d sources. Grey diamonds are ISO upper limits of the 
  $11.2$\,$\mu$m feature strength for Herbig Ae sources, grey `+' symbols are ISO detections. 
c2d sources are labelled as follows. 1: 
  \object{T\,Cha}; 2: \object{LkH$\alpha$\,330}; 3: \object{SR\,21 N}; 
  4: \object{HD\,135344}; 5: \object{RR\,Tau}; 6: \object{VV\,Ser}; 
  7: \object{HD\,101412}; 8: \object{HD\,98922}. The dotted grey line indicates 
  our typical 3$\sigma$ sensitivity limit of $2.5 \times 10^{-16}$ W m$^{-2}$ for 
  sources at $d = 150$ pc.}
  \label{fig:pah11_str_chi}
\end{figure}

\begin{figure}
  \centering
  \includegraphics[width=\columnwidth]{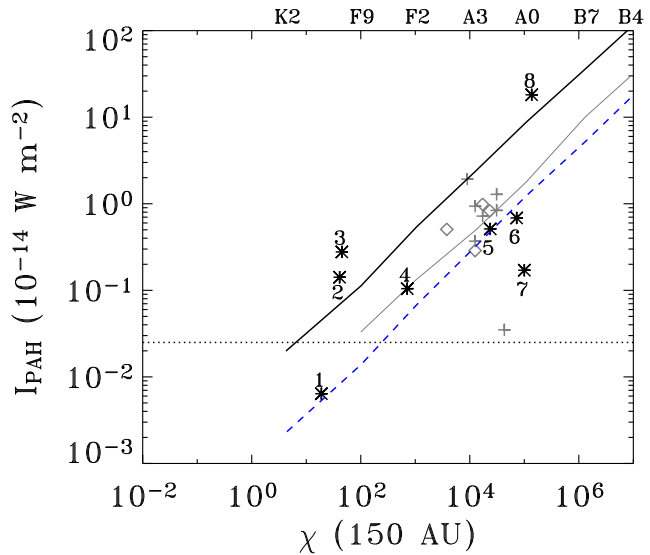}
  \caption{As Fig.~\ref{fig:pah11_str_chi}, but assuming a blackbody
  stellar spectrum and main-sequence stellar parameters. Dark solid
  line represents our template model, dashed line our
  model with a 10$\times$ lower PAH abundance. Light solid line represents
  the H04 disk model results for the $11.2$\,$\mu$m feature. }
  \label{fig:pah11_str_bb_chi}
\end{figure}

We have calculate $\chi$ for our models as presented in
Fig.~\ref{fig:pah11_str_temp} using Kurucz spectra and pre-main sequence
parameters. The results are presented in Fig.~\ref{fig:pah11_str_chi}. Moreover, we
have run our template model assuming a blackbody stellar spectrum and
main-sequence stellar parameters, with $T_{\mathrm{eff}}$ = 15000,
12300, 10500, 8500, 7000, 6000 and 5000 K, corresponding to spectral
types B4, B7, A0, A3, F2, F9 and K2, respectively
\citep{gra94}.  Fig.~\ref{fig:pah11_str_bb_chi} shows the
results for the template PAH abundance (same as H04, dark solid line)
and 10$\times$ lower (dashed line) and includes the model
prediction by H04 (light solid line). The observational data are included in
both figures, with $\chi$ computed from the spectral types in the same
way as described above for the pre-main sequence stars.

It is seen from Fig.~\ref{fig:pah11_str_chi} and Fig.~\ref{fig:pah11_str_bb_chi} that
for the same PAH abundance, our models predict factors of 4--5
stronger PAH features than the H04 models. In fact, our entire SED
(continuum {\em and} features) is stronger than in their models. This
means that our model disk absorbs more stellar radiation and hence
re-emits more radiation in the infrared, both in PAHs and in thermal
continuum.  A factor of two can be traced back to the different
assumption for the irradiative flux, whereas another $\sim$\,40\% is due
to the ``boosting'' effect in the 3-D versus 1+1-D approach in this
near-face-on geometry.  The remaining factor of $\sim$\,2 is ascribed to
the other differences described above in terms of disk structure,
geometry, treatment of dust grains, PAH destruction and PAH opacities. Another difference is that our template model assumes a mix of neutral and ionized PAHs whereas that of H04 assumes only neutral PAHs. A comparison of feature-over-continuum ratios, in which many of these effects drop out, would be instructive but H04 do not quote such values.

Comparison of Fig.~\ref{fig:pah11_str_chi} and Fig.~\ref{fig:pah11_str_bb_chi} shows
  that the 11.2 $\mu$m fluxes are lower if blackbody radiation is
  assumed with main sequence stellar parameters, especially for T
  Tauri stars (later than F8).  Both our blackbody models with
  10$\times$ lower abundance as well as the H04 models fit most of the
  ISO and Spitzer detections for HAeBe stars. Both models predict that
  for T Tauri stars of spectral type G5 and later, $I_\mathrm{PAH}$ to
  falls below our Spitzer detection limit for $d=150$ pc.  However,
  the H04 models underpredict the detected 11.2 $\mu$m strengths of
  some of the lowest luminosity sources in our sample, whereas our
  models with standard PAH abundance come much closer. In general, the
  PAH emission from T Tauri stars, when detected, appear better
  modeled with Kurucz spectra for pre-main sequence stars than with
  blackbody spectra for main sequence parameters. Overall, the main
  conclusions from our paper are not affected qualitatively if the H04
  models with blackbodies were used to interpret the
  data. Quantitatively, the threshold for detection shifts to earlier
  spectral types (G5 versus K5) and the inferred PAH abundances are
  factors of 4--5 higher.

Finally, we note that a multitude of tests of our code were
performed in addition to the tests that the {\tt RADMC} code already
has gone through \citep[e.g.][]{pas04}. We tested that our model
conserves luminosity; we compared the SED without PAHs to the SED
produced by the Chiang \& Goldreich-type model described in
\citet{dul01}; we estimated what the SED strength should be on the
basis of the disk geometric thickness; and we performed optically thin
models consisting purely of PAH grains and compared to the models of
\citet{li01}. All these tests confirmed the validity of the models.
\begin{table*}
\centering
\caption{List of differences between the model presented here and that of \citet{hab04}.}
\label{tbl:modelcomp}
\begin{tabular}{lll}
\hline
\hline
\noalign{\smallskip}
Model aspect  & This paper & H04 \\ 
\hline
\noalign{\smallskip}
Radiative transfer & Axisymmetric 3-D & 1+1-D \\
Disk structure        & Inner rim &  - \\
$\rho$($Z$)       & Parametrized & Self-consistent \\
Stellar spectrum    & Kurucz & BB \\
Stellar parameters    & PMS & MS \\
Stellar flux         & 100\% & 50\% \\
PAH evaporation  &  1/2  & yes \\
$T-$decoupling         & no      & yes \\
Optical-IR PAH opacities  & \citet{li01}, & \citet{li01},\\
 & \citet{mat05a} & \citet{des90}\\
\hline
\end{tabular}
\end{table*}

\Online
\section{Online Material}
\begin{figure*}
  \centering
  \includegraphics[width=17cm]{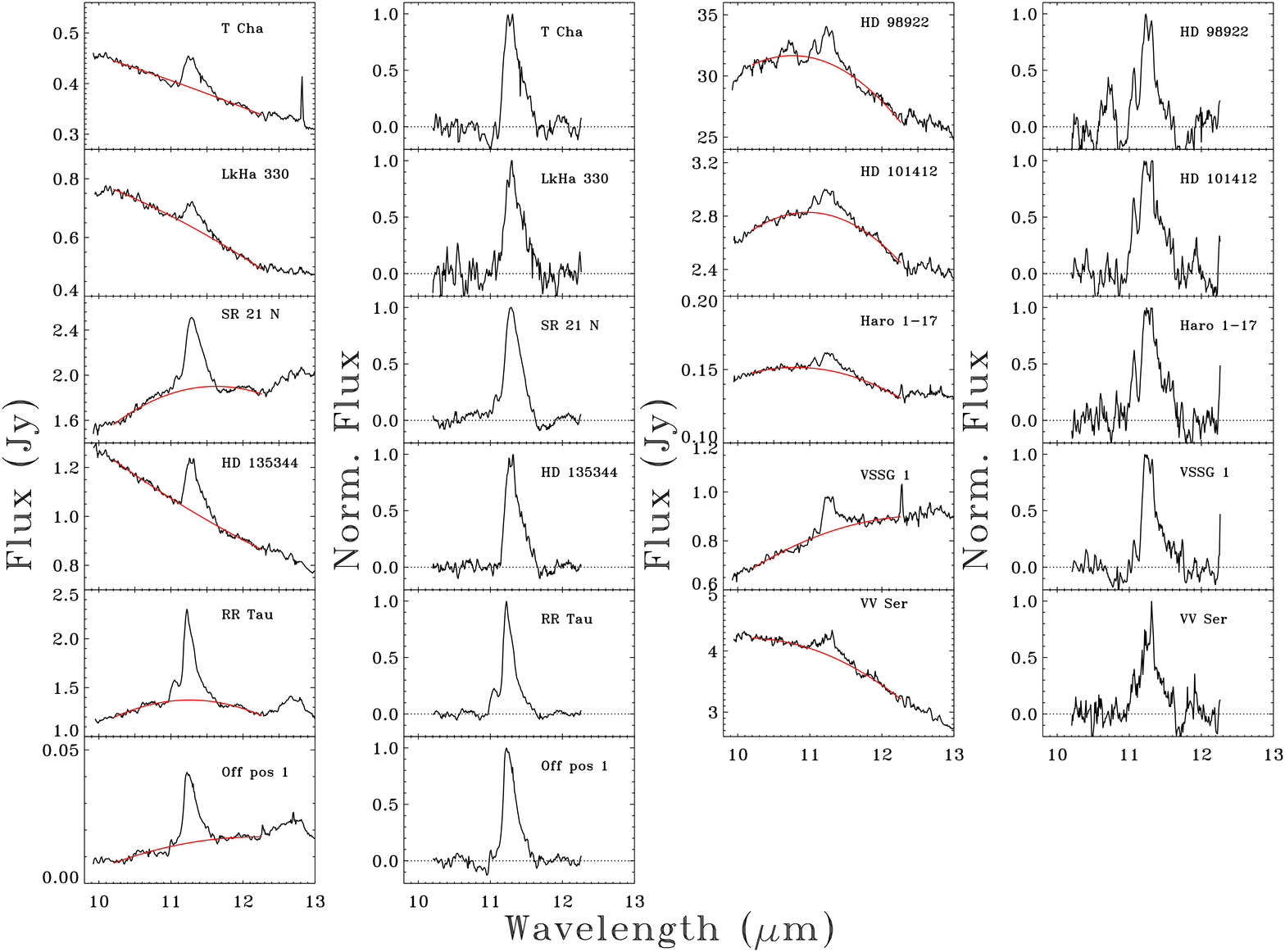}
 \caption{{\bf Left:} Blow-up of the Spitzer IRS high resolution spectra around the 11.2 $\mu$m PAH feature. A simple fit of the continuum flux below the PAH feature is plotted with a dotted line. {\bf Right:} Continuum subtracted spectra, normalised to the peak flux of the PAH feature.}
  \label{fig:pah11_sh_specfit}
\end{figure*}

\begin{figure*}
  \centering
  \includegraphics[width=17cm]{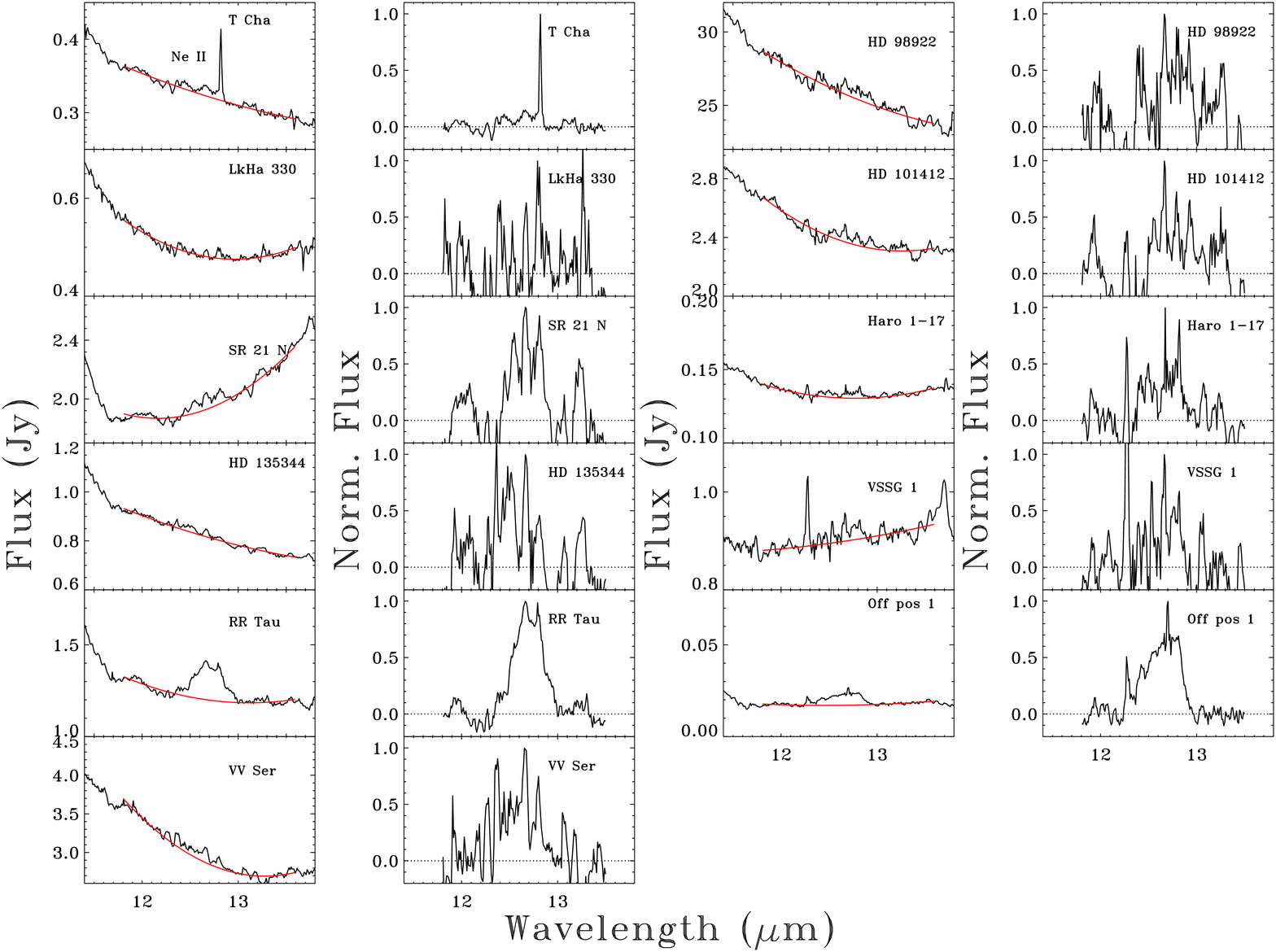}
  \caption{{\bf Left:} Blow-up of the Spitzer IRS high resolution spectra around the 
12.8 $\mu$m PAH feature. A simple fit of the continuum flux
  below the PAH feature is plotted with a dotted line. {\bf Right:} Continuum
  subtracted spectra, normalised to the peak flux of the PAH feature. }
  \label{fig:pah12_sh_specfit}
\end{figure*}

The Online Material contains Figs.\,\ref{fig:pah11_sh_specfit} and \ref{fig:pah12_sh_specfit}, presenting blow-ups of the $11.2$ and $12.8$ $\mu$m PAH features.
%
\end{document}